\documentclass[11pt]{article}

\usepackage{stolenstyle}

\usepackage{amsmath,epsf,amssymb,latexsym,amsthm,setspace,bbm,array,pifont}

\usepackage[matrix,arrow,frame,import,curve,color]{xy}

\DeclareFontFamily{U}{rsf}{}
\DeclareFontShape{U}{rsf}{m}{n}{
  <5> <6> rsfs5 <7> <8> <9> rsfs7 <10-> rsfs10}{}
\DeclareMathAlphabet\Scr{U}{rsf}{m}{n}

\def\CO#1#2{{[#1,#2]}}
\def\AC#1#2{{\{#1,#2\}}}

\def\rep#1{{{\boldsymbol{#1}}}}
\def\brep#1{{{\overline{\boldsymbol{#1}}}}}

\def\cDb{{\overline{\cD}}}

\def\GUL{\GU(1)_{\text{L}}}
\def\GUR{\GU(1)_{\text{R}}}

\def\C{{\mathbb C}}

\def\P{{\mathbb P}}
\def\R{{\mathbb R}}
\def\Z{{\mathbb Z}}

\def\Tr{\operatorname{Tr}}

\def\SO{\operatorname{SO}}

\def\GL{\operatorname{GL}}

\def\SU{\operatorname{SU}}
\def\GU{\operatorname{U{}}}

\def\GE{\operatorname{E}}

\def\so{\operatorname{\mathfrak{so}}}

\def\Lu{\operatorname{\mathfrak{u}}}
\def\Le{\operatorname{\mathfrak{e}}}

\def\p{\partial}

\def\la{\langle}
\def\ra{\rangle}

\def\ff#1#2{{\textstyle\frac{#1}{#2}}}

\def\cA{{\cal A}}

\def\cD{{\cal D}}

\def\cF{{\cal F}}

\def\cJ{{\cal J}}
\def\cK{{\cal K}}
\def\cL{{\cal L}}
\def\cM{{\cal M}}

\def\cO{{\cal O}}

\def\cU{{\cal U}}
\def\cV{{\cal V}}
\def\cW{{\cal W}}

\def\ep{{\epsilon}}

\newcommand\alphab{\overline{\alpha}}

\newcommand\thetab{\overline{\theta}}

\newcommand\taub{\overline{\tau}}

\newcommand\Sigmah{\widehat{\Sigma}}

\newcommand\Gammab{\overline{\Gamma}}

\newcommand\Phib{\overline{\Phi}}

\newcommand\cb{\overline{c}}

\newcommand\hb{\overline{h}}

\newcommand\jb{\overline{\jmath}}

\newcommand\wb{\overline{w}}

\newcommand\zb{\overline{z}}

\newcommand\Cb{\overline{C}}

\newcommand\Fb{\overline{F}}

\newcommand\Jb{\overline{J}}
\newcommand\Kb{\overline{K}}

\newcommand\Qb{\overline{Q}}

\newcommand\Wt{\widetilde{W}}
\newcommand\Xt{\widetilde{X}}
\newcommand\Yt{\widetilde{Y}}

\def\bJ{{{\boldsymbol{J}}}}

\def\bq{{{{\boldsymbol{q}}}}}
\def\bqb{{{\overline{\bq}}}}

\def\bpsi{{\boldsymbol{\psi}}}
\def\bpsib{{\overline{\bpsi}}}

\def\bj{{\boldsymbol{\jmath}}}
\def\bjb{{\overline{\bj}}}

\def\tint{{{\text{int}}}}

\def\GULIR{{{\GUL^{\text{IR}}}}}
\def\GURIR{{{\GUR^{\text{IR}}}}}

\def\preprint{ {{ MIFPA-14-17\\ NSF-KITP-14-043}}}

\title{Accidents in (0,2) Landau-Ginzburg theories}
\author[a,b] {Marco Bertolini,}
\author[c] {Ilarion V.~Melnikov,}
\author[a] {and M.~Ronen Plesser}
\affiliation[a]{Center for Geometry and Theoretical Physics, Box 90318 \\
Duke University, Durham, NC 27708-0318, USA}
\affiliation[b]{Kavli Institute for Theoretical Physics\\
University of California, Santa Barbara, CA 93106-4030, USA}
\affiliation[c]{ George P. and Cynthia W. Mitchell Institute for Fundamental Physics and Astronomy,
Texas A\&M University,
College Station, TX 77843, USA}
\emailAdd{mb266@phy.duke.edu}
\emailAdd{ilarion@physics.tamu.edu}
\emailAdd{plesser@cgtp.duke.edu}
\abstract{We study the role of accidental symmetries in two-dimensional (0,2) superconformal field theories obtained by RG flow from (0,2) Landau-Ginzburg theories.  These accidental symmetries are ubiquitous, and, unlike in the case of (2,2) theories, their identification is key to correctly identifying the IR fixed point and its properties.  We develop a number of tools that help to identify such accidental symmetries in the context of (0,2) Landau-Ginzburg models and provide a conjecture for a toric structure of the SCFT moduli space in a large class of models.  We also give a self-contained discussion of aspects of (0,2) conformal perturbation theory.}

\begin{document}

\maketitle

\section{Introduction}\label{s:intro}
The construction and classification of conformal field theories (CFTs) plays a key role in modern quantum field theory.  One approach to this problem is to solve the conformal bootstrap.\footnote{
This is difficult in practice and can be carried out analytically only in theories with enormously enlarged symmetries like the $W_N$ algebras of the minimal models~\cite{Belavin:1984vu}.  Recently progress has been made with numeric techniques,  for example in applications to the three-dimensional Ising model~\cite{El-Showk:2014dwa,ElShowk:2012ht}.}  Another approach that has proven useful is to study the low energy (or IR) limits of  renormalization group (RG) flows from known CFTs.  This has challenges of its own, since the IR dynamics often involves emergent degrees of freedom and interactions.

Nevertheless, as already indicated in the seminal work of~\cite{Zamolodchikov:1986db,Kastor:1988ef,Martinec:1988zu,Vafa:1988uu}, it is often possible to identify certain classes of operators and their OPEs and correlators of an IR CFT with corresponding objects in terms of the UV degrees of freedom.  This is especially useful when the UV theory is asymptotically free, since then perturbative computations can provide information about a non-trivial CFT without a notion of a weak coupling.  The identification of UV and IR data is simplified when some amount of supersymmetry is preserved along the RG flow:  SUSY constraints lead to well-known simplifications such as the relation between dimensions and R-charges of chiral operators and non-renormalization theorems.  For instance, in two dimensional theories with (2,2) SUSY these structures are responsible for many well-known phenomena such as mirror symmetry and the Calabi-Yau (CY) / Landau-Ginzburg (LG) correspondence~\cite{Hori:2003ds}, and the identification of UV and IR data is a key tool in exploration and exploitation of these two-dimensional gems.

Such techniques rely on the assumption that accidental symmetries that might emerge in the IR limit do not invalidate the identification of operators in the IR with their UV avatars.   This assumption is well-tested in (2,2) theories but is also often applied to theories with only (0,2) supersymmetry.  For instance, it is key to various gauged linear sigma model constructions of (0,2) CFTs corresponding to heterotic string vacua~\cite{Witten:1993yc,Distler:1993mk,Silverstein:1995re,Distler:1995mi}.  

In this note we show that the assumption cannot be taken for granted in (0,2) theories, and the resulting ``accidents'' have drastic consequences for the IR physics and the relation between UV parameters and IR data.  The examples we consider are (0,2) Landau-Ginzburg theories, and we identify a class of accidental symmetries of (0,2) LG RG flows by studying the space of F-term UV couplings modulo field redefinitions.  We find that these accidental symmetries significantly modify the analysis of the IR theory. 
For instance, the spectrum of chiral operators and even the IR central charge are in general modified.  This invalidates certain UV theories from giving good models for (0,2) SCFTs appropriate for a heterotic string vacuum---we examine an example taken from~\cite{Distler:1993mk}.

\subsubsection*{A classic (2,2) example}
To describe the challenges of (0,2) accidents more precisely, it is useful to review the successes of the (2,2) theories.
Consider the quintic (2,2) LG model with chiral superpotential
\begin{align*}
W = \alpha_0 X_0^5 + \alpha_1 X_1^5+\cdots +\alpha_4 X_4^5 -5 \alpha_5 X_0 X_1 \cdots X_4~.
\end{align*}
Here the $X_i$ are (2,2) chiral superfields and the $\alpha_a$ are complex parameters.  This theory flows to a $c=\cb = 9$ (2,2) SCFT.\footnote{A $\Z_5$ orbifold of the theory describes a (2,2) non-linear sigma model with target space the quintic CY hypersurface in $\P^4$ at a special value of the complexified K\"ahler parameter.  A complex structure parameter of the geometry is then related to the LG parameter $\psi$.}  The complex parameter
\begin{align*}
\psi = {\alpha_5^5}({\alpha_0\alpha_1\cdots\alpha_4})^{-1}~
\end{align*}
is invariant under $\C^\ast$ rescalings of the chiral superfields $X_i$ and labels a one-parameter family of IR CFTs.
At generic values of $\psi$ the IR fixed point is a well-behaved CFT, and small changes in $\psi$ correspond to small marginal deformations of the CFT, where ``small'' refers to the distance in the Zamolodchikov metric.  At special values of $\psi$ the CFT can become singular.  For instance, $\psi =1 $ is a finite distance singularity ---the analogue of a conifold point.  This can be detected in the UV description:  the theory develops a family of supersymmetric vacua with $X_i = \text{const}$., and these signal a non-compact CFT:  a theory with a continuum spectrum of conformal dimensions.  Another point, $\psi = \infty$ is an infinite-distance singularity.

The quotient of the UV parameter space by field redefinitions is a complicated object~\cite{Aspinwall:1993rj,Cox:2000vi} with singularities and non-separated points.  For instance, we can take the limit $\alpha_0 \to 0$ and $\alpha_5 \to 0$ while keeping $\psi$ constant so as to obtain a product of four minimal models coupled to a free chiral superfield $X_0$, with $c=\cb = 3 (3 + \ff{2}{5})$.  Fortunately, all such bad points are singular CFTs. The bad points corresponding to various infinite distance singularities and ``wrong'' central charges are easily identified in terms of the UV data:  they all correspond to singular superpotentials with a continuum of supersymmetric vacua and therefore a continuum of states in the IR CFT.  Away from such points the quotient is sensible and describes (2,2) marginal deformations of the IR theory.  

Similar considerations apply to more general (2,2) gauged linear sigma models, and with the parametrization of the smooth CFTs in terms of the UV parameters  in hand, localization and topological field theory techniques can be used to compute certain correlators and chiral spectra in the CFT in terms of the weakly coupled UV Lagrangian.

\subsubsection*{(0,2) challenges}
As we will show in section~\ref{s:accidents}, the situation is more delicate in (0,2) theories,  even in the relatively simple class of LG models (we review these in section~\ref{s:review}).  The essential difference is that we lack the simple diagnostic we had for a ``bad'' point in (2,2) theories.  It is not sufficient to exclude UV parameters that lead to flat directions in the potential, and the identification of UV parameters with marginal deformations of the CFT requires (at least) a study of loci with enhanced symmetry.  Unlike in (2,2) examples, accidental symmetries can emerge for non-singular UV potential, thereby complicating the description of IR physics in terms of the UV data.  Unlike in (2,2) theories a family of smooth UV potentials with each potential preserving the same R-symmetry  along the RG flow need not correspond to a family of CFTs related by truly marginal deformations.

Fortunately, at least in (0,2) LG models it appears that we have enough control to identify accidental symmetries and special loci in the parameter space by generalizing the (2,2) paradigm of parametrizing the IR fixed points by the space of UV parameters modulo field redefinitions.  This uncovers a rich structure of (0,2) RG flows and of the space of marginal deformations of (0,2) fixed points and will undoubtedly play a role in quantitatively descriptions of (0,2) moduli spaces.

Once we have identified a (0,2) LG theory with some particular IR fixed point, it is useful to develop the correspondence between deformations of the UV Lagrangian and (0,2) conformal perturbation theory.  In section~\ref{s:CPT} we describe some properties of (0,2) conformal perturbation theory independent of any embedding of the CFT in a critical heterotic string.  This section can be read independently from the rest of the paper.  We use these observations in section~\ref{s:toric} to describe a conjecture for the global structure of the moduli space of (0,2) SCFTs with expected central charge in terms of the UV data for what we term \textit{plain} (0,2) LG models.

\acknowledgments IVM would like to thank the Enrico Fermi Institute for hospitality while this work was being completed and especially David Kutasov for discussions.  This work is supported in part by the NSF Focused Research Grant DMS-1159404  and Texas A\&M (IVM), by NSF Grant PHY-1217109 (MB and MRP) and NSF Grant PHY11-25915 (MB).  MRP thanks the Aspen Center for Physics (and NSF Grant 1066293) for hospitality during the initial stage of this work.   We thank the organizers of the workshop on Heterotic Strings and (0,2) QFT at Texas A\&M  University in May, 2014 for hosting MB and MRP while some final steps were completed, and for allowing MRP to present a preliminary version.   We also thank Paul Aspinwall and Eric Sharpe for some useful conversations, and IVM and MPR thank Ido Adam for discussions and collaboration on aspects of (0,2) conformal perturbation theory.

\section{A glance at (0,2) Landau-Ginzburg theories} \label{s:review}
We begin with a quick review of (0,2) LG theories following~\cite{Distler:1993mk,Kawai:1993jk,Melnikov:2009nh}.\footnote{Our superspace conventions are those of~\cite{Melnikov:2009nh}.}  
We work in Euclidean signature and a (0,2) superspace with coordinates $(z,\zb,\theta,\thetab)$. 
The UV theory consists of $n$ bosonic chiral multiplets $\Phi_i = \phi_i + \ldots,$ and $N$ fermionic chiral multiplets $\Gamma^A = \gamma^A+\ldots$, as well as their conjugate anti-chiral multiplets.  These are given a free kinetic term
and a (0,2) SUSY potential term as interactions:
\begin{align}
\cL_{\tint} &= \int \! d\theta ~\cW + \text{h.c.}~,&
\cW = m_0\textstyle\sum_A \Gamma^A J_A(\Phi)~,
\end{align}
where $m_0$ has mass dimension $1$ and the $J_A$ are polynomial in the $\Phi_i$.   This is the simplest example of a (0,2) SUSY asymptotically free theory: for energies $E \gg m_0$ the theory is well-described by the set of free fields.  Conversely, when $E \ll m_0$ the interactions become important and lead to non-trivial IR dynamics that depend on $n,N$, as well as the choice of ideal $\bJ = \la J_1,\ldots,J_N\ra \subset \C[\Phi_1,\ldots,\Phi_n]$.  What can we say about the IR limit of this theory?

A basic constraint comes from the gravitational anomaly.  In the UV the central charges are easy to determine:  each $\Phi$ multiplet contains a complex boson and a right-moving Weyl fermion, while each $\Gamma$ contains a left-moving Weyl fermion and an auxiliary field.  Hence, we have $c_{UV} = 2n +N$, $\cb_{UV}  = 3n$.
The RG flow induced by $\cW$ will decrease the central charges, but since it is Lorentz-invariant, it will preserve the difference $c-\cb = N-n$.

Another basic property of the theory is the set of global symmetries.  The free theory has a large global symmetry that commutes with (0,2) SUSY: namely, $\GU(n)\times\GU(N)$ rotations of the chiral superfields.  In addition we have the R-symmetry that rotates $\theta$ and leaves the lowest components of the superfields $\phi_i$ and $\gamma^A$ invariant.  The interactions break these symmetries.  For completely generic $J_A$ the remaining symmetry just $\GUR^0$ -- an R-symmetry that assigns charge $+1$ to $\theta$ and $\gamma^A$ and charge $0$ to $\phi_i$.

\subsubsection*{Properties of the superpotential}
A key feature of (0,2) LG theories is that the holomorphic superpotential obeys the same non-renormalization properties as the, perhaps more familiar N=1 d=4 Wess-Zumino model's superpotential.  The kinetic term, on the other hand, is a full superspace derivative and will receive complicated corrections along the RG flow.  However, just as in (2,2) theories, we expect these corrections to be irrelevant provided that the fields $\Phi$ and $\Gamma$ all acquire non-trivial scaling dimensions.  In order to relate these scaling dimensions to properties of the UV theory, we will assume that the interactions preserve an additional global $\GU(1)$ symmetry, which we will call $\GUL$, under which $\Gamma^A$ have charges $Q_A$, while the $\Phi_i$ carry charges $q_i$.  This will be the case if and only if the ideal is quasi-homogeneous, i.e.
\begin{align}
J_A( t^{q_i} \Phi_i) = t^{-Q_A} J_A (\Phi)~
\end{align}
for all $t\in \C^\ast$.   We will demand that the ideal is zero-dimensional, i.e. $J_A(\Phi) = 0$ for all $A$ if and only if $\Phi = 0$.  If it is not then the theory necessarily has a non-compact set of supersymmetric vacua labeled by vevs of the bosonic fields.  We will call such superpotentials singular.  We are interested in ``compact'' CFTs and exclude this possibility.\footnote{A CFT is compact if its spectrum is such that for every fixed real $\Delta$ there is a finite number of fields with dimension less than $\Delta$.}  

Another important property of the superpotential is that typically some of its parameters are not, in fact, F-terms.  To see this, consider a perturbation of the form $\delta \cW = \sum_A \Gamma^A \delta J_A$  around a theory with $\cW_0 = \sum_A\Gamma^A J_A$.  In the undeformed theory, if we assume canonical kinetic terms, the equations of motion read
\begin{align}
\cDb \Gammab^A &= J_A(\Phi)~,& \cDb \p_z \Phib_i &= \sum_{A} \Gamma^A\frac{\p J_A}{\p \Phi_i}~,
\end{align}
where $\cDb$ is the antichiral superspace derivative $\cDb = \p_{\thetab} +\theta\p_{\zb}$.
A more general kinetic term leads to more complicated expressions under the $\cDb$ derivative of the left-hand sides of the equations.\footnote{In a NLSM such total derivatives are more subtle than in this LG setting, as they are usually only sensible patch by patch in target space.  Indeed, marginal deformations of NLSMs are such F-terms that cannot be globally recast as D-terms~\cite{Beasley:2004ys,Melnikov:2011ez}.}      Hence, a first order deformation of $\cW$ of the form
\begin{align}
\delta J_A = \sum_{B} M^B_A(\Phi) J_B(\Phi) + \sum_i \frac{\p J_A}{\p \Phi_i} F_i(\Phi) 
\end{align}
is equivalent up to equations  of motion to a D-term deformation. 

The LG assumption that the D-terms are irrelevant along the flow implies that any two UV theories with superpotentials related by a holomorphic field redefinition lead to the same IR fixed point.  Hence, any two UV superpotentials that are related by a holomorphic field redefinition belong to the same universality class.

\subsubsection*{(2,2) LG theories}
The (0,2) theory will have an enhanced left-moving SUSY when $N=n$, so that in the free limit we can combine $(\Gamma^i,\Phi_i)$ into (2,2) chiral multiplets $X^i$, and when $J_i = \p W/\p \Phi_i$ for some potential $W$.  In that case, we can rewrite the theory in a manifestly (2,2) supersymmetric fashion with a chiral superpotential $W(X)$.  The quasi-homogeneity conditions set $Q_i = q_i -1$, and the resulting central charge is given by the famous
\begin{align}
\cb = 3 \sum_i (1-2 q_i)~.
\end{align}

\subsubsection*{IR consequences of UV symmetries}
Returning to the more general (0,2) setting, if we assume that $\GUR^0$ and $\GUL$ are the only symmetries along the whole RG flow, then we can determine the linear combination of charges corresponding to the IR R-symmetry $\GUR^{\text{IR}}$ as well as those of a left-moving $\GULIR$~.  The charges of the latter are fixed up to normalization by the quasi-homogeneity condition, and the normalization is fixed by
\begin{align}
\label{eq:normfix}
-\sum_A Q_A -\sum_i q_i = \sum_A Q_A^2 - \sum_i q_i^2~.
\end{align}
This ensures that $\GULIR$ and $\GURIR$ have no mixed anomalies and become, respectively, left-moving and right-moving Kac-Moody symmetries in the IR theory.  The central charge is determined from the two-point function of the $\GURIR$ current.  The result is
\begin{align}
\label{eq:boringcharges}
\cb &= 3 (r + n-N)~, & r = -\sum_A Q_A -\sum_i q_i~.
\end{align}
By studying the cohomology of the supercharge $\Qb$ of the theory, we can also describe chiral operators and their charges.
More details can be found in~\cite{Melnikov:2009nh}, but for our purposes it will be sufficient to note the charges
and corresponding dimensions of $\phi_i$ and $\gamma^A$.  Denoting the $\GULIR$ and $\GURIR$ charges by, respectively, $\bq$ and $\bqb$, we have
\begin{align*}
\xymatrix@R=1mm@C=1cm{
~		&\phi_i		&\gamma^A \\
\bq		&q_i			&Q_A	\\
\bqb		&q_i			&1+Q_A	\\
h		& \ff{q_i}{2}	& \ff{2+Q_A}{2}	\\
\hb		&\ff{q_i}{2}	& \ff{1+Q_A}{2} 
}
\end{align*}
Since these are chiral operators, the right-moving weights are determined in the usual fashion $\hb = \bqb/2$, and the left-moving weights are fixed since RG flow preserves the spin of the operators.

This structure determines many properties of the IR theory such as the elliptic genus~\cite{Kawai:1993jk} and the topological heterotic ring~\cite{Melnikov:2009nh}.  As for (2,2) theories, there is also a simple prescription for using orbifolds of such (0,2) LG theories to build spacetime SUSY heterotic string vacua~\cite{Distler:1993mk,Blumenhagen:1995ew}.  For instance, the elliptic genus is given by~\cite{Kawai:1993jk}
\begin{align}
Z(\tau,z) &= \Tr_{RR} (-)^F y^{J^{\text{IR}}_{L0}} e^{2\pi i\tau H_{L}} e^{-2\pi i\taub H_{R}} \nonumber\\
& = i^{N-n} e^{2\pi i\tau (N-n)/12} y^{-r/2}\left[\chi(y) + O(e^{2\pi i \tau}) \right]~,
\end{align}
where  $y=e^{2\pi i z}$, and
\begin{align}
\chi(y) = \left. \frac{\prod_A (1-y^{-Q_A})}{\prod_i (1- y^{q_i})} \right|_{y^\text{integer}}~.
\end{align}
The remaining $\tau$-dependent terms are determined by modular properties of $Z(\tau,z)$.

\subsubsection*{Enhanced symmetry and $c$-extremization}
For special values of the superpotential the UV theory will acquire enhanced symmetries that commute with the (0,2) SUSY algebra. In two dimensions these cannot be spontaneously broken, and, as in four dimensions, the abelian component $\GU(1)^M$ can mix with $\GUR^0$ and $\GUL$ symmetries.  Fortunately, as in the four-dimensional case we can still find candidate $\GULIR$ and $\GURIR$ symmetries by applying the analogue of $a$-maximization~\cite{Intriligator:2003jj} known as $c$-extremization~\cite{Benini:2012cz}.   We can summarize the results of~\cite{Benini:2012cz} as follows.  Let $\cJ_0$ denote the $\GUR^0$ R-symmetry current, and  let $\cJ_\alpha$, $\alpha = 1,\ldots, M$ be the currents for $\GU(1)^M$.   Assuming that the correct $\GURIR$ symmetry is a linear combination of $\cJ_0$ and the $\cJ_\alpha$,~\cite{Benini:2012cz} construct the trial current and trial central charge
\begin{align}
\cJ &= \cJ_0 + \sum_\alpha t^\alpha \cJ_\alpha~, &
\ff{1}{3}\Cb & = n-N + 2\sum_{\alpha} t_\alpha K^\alpha + \sum_{\alpha,\beta} t^\alpha t^\beta L_{\alpha\beta}~,
\end{align}
where
\begin{align}\label{eq:knl}
K^\alpha &= -\sum_A Q^\alpha_A - \sum_i q_i^\alpha~,&
L^{\alpha\beta} & =   \sum_i q_i^\alpha q_i^\beta-\sum_A Q^\alpha_A Q_A^\beta~.
\end{align}
Here $Q^\alpha_A$ and $q^\alpha_i$ denote the $\GU(1)^\alpha$ charges of $\Gamma^A$ and $\Phi_i$, respectively.  The $\GURIR$ is then identified by extremizing $\Cb$ with respect to $t_\alpha$, leading to $\GULIR$ charges
\begin{align}
q_i & = \sum_\alpha q_i^\alpha t_{\alpha\ast}~,&
Q_A & = \sum_\alpha Q_A^\alpha t_{\alpha\ast}~,
\end{align}
where $t_\alpha = t_{\alpha\ast}$ is the extremum point.  The central charge is then also fixed as $\cb = \Cb(t_\ast)$.

The symmetric form $L$ has a real spectrum, and the sign of an eigenvalue has the following significance in the IR theory. We decompose the UV currents according to the sign of the eigenvalues as $\cJ_\alpha \to \{\cJ^+, \cJ^0,\cJ^-\}$.  If we assume that there are no accidental symmetries in the IR, then unitarity of the SCFT implies that in the IR the $\cJ^+$ currents must correspond to right-moving Kac-Moody (KM) currents and the $\cJ^-$ must flow to left-moving KM currents.  Finally, the $\cJ^0$ must decouple from the SCFT degrees of freedom.  The last point has two consequences:  on one hand, we should treat a theory with $\ker L \neq 0$ with some care; on the other hand, if we can be certain that the IR limit is nevertheless a unitary CFT, we can without loss of generality restrict to symmetries orthogonal to $\ker L$.

In typical examples of (0,2) LG theories $L$ is negative definite; we do not know of a non-singular model where $L$ has a positive eigenvalue.  In fact, as far as the extremization procedure goes, symmetries corresponding to the positive eigenspace of $L$ cannot be broken in the SCFT.  More precisely, a UV deformation away from an RG trajectory with a ``positive'' symmetry is irrelevant --- in the IR the ``positive'' symmetry will be restored.  To understand this, we consider the change in the extremized central charge upon breaking a symmetry.  Assuming $\ker L=0$, the extremum central charge is
\begin{align}
\ff{1}{3} \Cb_{0} = n-N - K^T L^{-1} K~.
\end{align}
Now suppose we change parameters so that some of the symmetries are broken.  We can characterize the unbroken symmetries by a a vector $v^\alpha$, so that the unbroken symmetries satisfy $t^T v = 0$.  The modified extremization is then easily carried out with the aid of a Lagrange multiplier $s$:
\begin{align}
\ff{1}{3} \Cb_v(t,s) = n-N + 2 t^T K + t^T L t + 2s t^T v~.
\end{align}
Extremizing with respect to $t$ leads to
\begin{align}
\ff{1}{3} \Cb_v(s) = \ff{1}{3}\Cb_0 - 2 s v^T L^{-1} K - s^2 v^T L^{-1} v~.
\end{align}
This may be extremized for $s$ if and only if $v^T L^{-1} v \neq 0$, in which case we obtain
\begin{align}
\label{eq:decrease}
\ff{1}{3} \Cb_v = \ff{1}{3} \Cb_0 + \frac{ (v^T L^{-1} K)^2}{v^T L^{-1} v}~.
\end{align}
The first observation is that the deformation changes the IR central charge if and only if the original symmetry, with charges determined from $t_\ast = -L^{-1} K$ is broken.  Next, we see that if in addition $v$ belongs to the positive eigenspace of $L$, then the central charge of the deformed theory is strictly greater than that of the undeformed theory --- this means the deformation must be irrelevant in the IR, and we expect the deformed theory to flow to the original undeformed fixed point.  Once we eliminate these irrelevant deformations from the parameter space, the symmetries corresponding to the positive eigenvalues of $L$ are never broken, and we can restrict to $v$ in the negative eigenspace of $L$.

We stress that in all examples we considered $L$ is negative definite.  In that case~(\ref{eq:decrease}) shows that when a deformation breaks a symmetry the central charge changes if and only if the deformation breaks the R-symmetry, and whenever that happens the central charge decreases.

\subsubsection*{Constraints on UV data}
The structure relating UV and IR physics sketched above assumes that for a given set of charges $(q_i, Q_A)$ there exists a non-singular potential with a zero-dimensional ideal $\bJ$ and of course that $\GUL$ and $\GUR^{0}$ are the only symmetries all along the RG flow.  Both of these are non-trivial assumptions.  

It is an open problem to classify all sets of charges consistent with~(\ref{eq:boringcharges}) and some fixed $\cb$ that can be realized by a non-singular $\bJ$.\footnote{This should be contrasted with (2,2) LG models, for which such a classification exists~\cite{Kreuzer:1992np,Klemm:1992bx} and yields a finite set of quasi-homogeneous potentials at fixed central charge.}  Demanding that $\chi(y)$ is a polynomial rules out many choices of charges, but while being a necessary condition, it is not sufficient to show that there exists a zero-dimensional $\bJ$ realizing the charge assignment.

The second assumption, which amounts to the statement that there are no accidental symmetries in the IR, also leads to some necessary conditions.  For instance, just as in N=1 $d=4$ SQCD~\cite{Seiberg:1994pq}, violation of unitarity bounds on the charges can indicate an inconsistency in the assumption.  In particular, we have the unitarity bounds
\begin{align}
0<q_i &\le \cb/3~,& 0 &< (1+Q_A) &  \sum_A (1+Q_A) &\le \cb/3~.
\end{align}
These arise by demanding that $\phi_i$, $\gamma^A$, and $\prod_A \gamma^A$ are chiral primary operators of a unitary N=2 superconformal algebra.  The latter is particularly strong and eliminates many possible candidate charges.\footnote{In the (2,2) case this translates to  the known bound $\sum_i q_i \le n/3$~\cite{Kreuzer:1992bi}.} While these criteria are important and will certainly play a role in any attempt to classify (0,2) LG theories, they are not sufficient to rule out accidents.

\section{Accidents}  \label{s:accidents}
Having reviewed the basic structure of (0,2) LG theories, we will now study it in a few examples that will illustrate some of the subtleties in their analysis.

\subsection{Accidents in (2,2) Landau-Ginzburg orbifolds}
There are two familiar examples of accidental symmetries in (2,2) flows.  Consider LG orbifolds with potentials
\begin{align}
W_3 &= X_1^3 + X_2^3 + X_3^3 - \psi X_1 X_2 X_3~, &
W_4 & =X_1^4 + X_2^4 + X_3^4 +X_4^4 - \psi X_1 X_2 X_3 X_4~,
\end{align}
For $W_3$ ($W_4$) we take the orbifold by $\Z_3 \subset \GUR$ ($\Z_4 \subset \GUR$).  The endpoint of the flow in each case has accidental symmetries.  In the case of $W_3$, which is a special point in the moduli space of a (2,2) compactification on $T^2$, there is an accidental N=2 Kac-Moody $\GU(1)$ algebra for both left and right movers, corresponding to the isometries of the torus.  In the case of $W_4$ the IR theory is actually a (4,4) SCFT, and there are additional currents that enhance $\GUL\times\GUR$ to $\SU(2)_{\text{L}}\times\SU(2)_{\text{R}}$.  Of course this is the case for any Landau-Ginzburg orbifold (or more generally linear sigma model) that corresponds to a locus in the moduli space of $T^2$ or K3 compactification.

\subsection{A contrived (2,2) example}
Consider a (2,2) LG theory with
\begin{align}
W = X^3 + Y^4~.
\end{align}
There is a unique assignment of R-charge $\bqb(X) = 1/3$ and $\bqb(Y) = 1/4$, and the IR fixed point is the $\GE_6$  minimal model.  On the other hand, we can make a field redefinition $\Xt = X-Y$ and $\Yt = Y$.  This is certainly non-singular and leads to a superpotential
\begin{align}
\Wt =  \Xt^3 +3 \Xt^2 \Yt + 3 \Xt \Yt^2 + \Yt^3 + \Yt^4~. 
\end{align}
If we also perform the field redefinition in the kinetic terms, we have of course done nothing; however, if we assume the D-terms are indeed irrelevant, then taking standard kinetic terms  and either $W$ or $\Wt$ interactions should lead to the same IR fixed point.  Unlike the original theory, the $\Wt$ theory has no manifest R-symmetry along the flow --- the symmetry emerges accidentally in the IR.

The example is very contrived, but it illustrates the basic issue:  field redefinitions can obscure the UV fields that should be identified with IR operators of some fixed scaling dimension.  As we show in~\ref{ss:22sym}, if we restrict to (2,2) theories with a quasi-homogeneous potential, this ambiguity turns out to be harmless.  As the next example shows, in (0,2) theories this is not the case.

\subsection{A simple (0,2) example} \label{ss:devilmodel}
Consider a theory with $N=3$, $n=2$ and superpotential
\begin{align}
\label{eq:devilmodel}
\cW_0 = \begin{pmatrix} \Gamma^1 & \Gamma^2 & \Gamma^3 \end{pmatrix} 
\begin{pmatrix} 
\alpha_{11} & \alpha_{12} & \alpha_{13} \\ 
\alpha_{21} & \alpha_{22} & \alpha_{23} \\ 
\alpha_{31} & \alpha_{32} & \alpha_{33}
\end{pmatrix}
\begin{pmatrix} \Phi_1^6 \\ \Phi_2^2 \\ \Phi_1^3 \Phi_2 \end{pmatrix}~. 
\end{align}
For generic values of the $9$ parameters $\alpha$ the potential preserves a unique $\GUL$ symmetry, and normalizing the charges as in~(\ref{eq:normfix}) leads to $r=2$, $\cb = 3$, and charge assignments
\begin{align}
\label{eq:devilcharges}
\xymatrix@R=0.0mm@C=1mm{~ & \Phi_1	& \Phi_2	&\Gamma^{1,2,3} \\
\bq	& \ff{1}{7} & \ff{3}{7}	& -\ff{6}{7}	  \\
\bqb	& \ff{1}{7} & \ff{3}{7}	& \ff{1}{7}
}
\end{align}
To obtain a description of the parameter space of the IR theory we consider the $\alpha$ modulo field redefinitions consistent with (0,2) SUSY and the $\GUL$ symmetry:
\begin{align}
\Gamma^A &\to \sum_B\Gamma^B M_{B}^A~,&
\Phi_1 & \to x \Phi_1~,&
\Phi_2 &\to y \Phi_2 + z \Phi_1^3~.
\end{align}
These transformations are invertible if and only if $M \in\GL(3,\C)$ and $x, y\in \C^\ast$.  The induced action on the $\Phi$ monomials is then
\begin{align}
\begin{pmatrix} \Phi_1^6 \\ \Phi_2^2 \\ \Phi_1^3 \Phi_2 \end{pmatrix} & \to
S \begin{pmatrix} \Phi_1^6 \\ \Phi_2^2 \\ \Phi_1^3 \Phi_2 \end{pmatrix}~,&
S= \begin{pmatrix} x^6 & 0 & 0 \\ x^3 z & x^3 y & 0 \\ z^2 & 2yz& y^2  \end{pmatrix}~,
\end{align}
and hence the action on the parameters $\alpha$ is $\alpha \mapsto M \alpha S$.

A bit of algebra shows that every non-singular ideal $\bJ$ described by $\alpha$ is equivalent by a field redefinition to one of three superpotentials:
\begin{align}
\cW_1 & = \Gamma^1 \Phi_1^6 + \Gamma^2 \Phi_2^2 + \Gamma^3 \Phi_1^3 \Phi_2~,\nonumber\\
\cW_2 & = \Gamma^1 (\Phi_1^6+\Phi_2^2) + \Gamma^2 \Phi_1^3 \Phi_2~,\nonumber\\
\cW_3 & = \Gamma^1 \Phi_1^6 + \Gamma^2 \Phi_2^2~.
\end{align}
The UV parameter space is stratified to three points, and we consider each in turn.
\begin{enumerate}
\item $\cW_1$ has a $\GU(1)^2$ global symmetry that acts independently on $\Phi_1$ and $\Phi_2$; extremization picks out the following charges.
\begin{align*}
\xymatrix@R=0.0mm@C=5mm{
~ 	& \theta		&\Phi_1		&\Phi_2		&\Gamma^1 	&\Gamma^2	&\Gamma^3 	&&\\
\bq	& 0			& \ff{26}{167}	&\ff{64}{167}	&-\ff{156}{167}	&-\ff{128}{167}	&-\ff{142}{167} && \cb = 3\left( 1 + \ff{2}{167} \right)\\
\bqb	& 1			&\ff{26}{167}	&\ff{64}{167}	&\ff{11}{167}	&\ff{39}{167}	&\ff{25}{167} 	&&
}
\end{align*}
\item $\cW_2$ has a free $\Gamma^3$ multiplet.  The interacting part of the theory has no extra global symmetries and $\GULIR\times\GURIR$ charges 
\begin{align*}
\xymatrix@R=0.0mm@C=5mm{
~ 	& \theta		&\Phi_1		&\Phi_2		&\Gamma^1 	&\Gamma^2	&&	\\
\bq	& 0			& \ff{4}{31}	&\ff{12}{31}	&-\ff{24}{31}	&-\ff{24}{31}	&&	\cb = 3 \left( 1 + \ff{1}{31} \right)~. \\
\bqb	& 1			&\ff{4}{31}		&\ff{12}{31}	&\ff{7}{31}		&\ff{7}{31}		&&	 
}
\end{align*}
\item $\cW_3$ has a free $\Gamma^3$ multiplet, and the interacting part of the theory is a product of (2,2) minimal models with (2,2) superpotential $W  = X_1^7 + X_2^3$ and charges
\begin{align*}
\xymatrix@R=0.0mm@C=5mm{
~ 	& \theta		&\Phi_1		&\Phi_2		&\Gamma^1 	&\Gamma^2	&&	\\
\bq	& 0			&\ff{1}{7}		&\ff{1}{3}		&-\ff{6}{7}		&-\ff{2}{3}	 	&& \cb = 3\left(1+ \ff{1}{21}\right). \\
\bqb	& 1			&\ff{1}{7}		&\ff{1}{3}		&\ff{1}{7}		&\ff{1}{3}	 	&&
}
\end{align*}
\end{enumerate}
If we assume that there are no accidental symmetries for the $\cW_1$, $\cW_2$ and $\cW_3$ theories, we obtain a consistent picture of the RG flows starting with the UV theory in~(\ref{eq:devilmodel}).  There are three basins of attraction; each has a central charge $\cb > 3$, a set of charges consistent with unitarity bounds and no marginal deformations.  Moreover, we can construct interpolating RG flows $\cW_3 \to \cW_2 \to \cW_1$ by adding relevant deformations to the superpotentials.  However, $\cW_1$ has no $\GUL$-invariant relevant deformations that make it flow to a putative $\cb = 3$ theory described by $\cW_0$.

We conclude that (0,2) LG RG flows have accidental symmetries, and identifying these is key in order to correctly pinpoint even basic properties of the IR theory.  For instance, we see in the example at hand that no point in the UV parameter space leads to an IR theory with $\cb = 3$ and $r=2$.

\subsection{Puzzles from enhanced symmetries}
There are two questions that probably occur to our erudite reader.  First, what's the big deal?  One has to take account of field redefinitions when discussing the parameter space of a theory, and it seems that all we learned here is that the parameter space is smaller than one may have naively thought.  Second, is it not perverse to discover some accidental symmetries associated to $\cW_{1,2,3}$ versus $\cW_0$ but then blithely assume that $\cW_{1,2,3}$ do not themselves suffer from accidents? 

There is a pragmatic answer to the second question: we assume there are no accidents unless we are able to identify some paradox in the putative description of the IR physics in terms of the UV parameters.  In our example we find such a paradox: while a generic $\cW$ has a unique global symmetry in the UV, there are special points with enhanced symmetries and a central charge that exceeds the putative $\cb = 3$ of the generic $\cW$!   Once we take into account the accidental symmetries, we discover that the enhanced symmetries are unavoidable, and there is no $\cb = 3$ theory that can be reached within the parameter space of these UV theories.  It is also easy to construct paradoxical examples that would violate unitarity bounds unless one takes accidents into account~\cite{Melnikov:2009nh}.

The answer to the first question is contained in this pragmatic perspective.  The ``big deal'' is that in the examples with which we are most familiar, namely the (2,2) LG theories, one never encounters these enhanced symmetry puzzles:  although there are plenty of points with enhanced symmetries, these never mix with $\GURIR$, and the central charge does not jump for any choice of non-singular (2,2) superpotential.  We discuss this in detail in the next section.

\subsection{Enhanced symmetries of (2,2) LG theories} \label{ss:22sym}
Consider a (2,2) LG theory with a quasi-homogeneous (2,2) superpotential $W(X)$ obeying $W(t^{q_i} X_i) = t W(X)$.  Unless $W$ satisfies an independent quasi-homogeneity condition, the (2,2) R-symmetries are fixed uniquely, giving charge $\bqb = q_i$ to $\Phi_i$ and $\Gamma^i$, where $(\Phi_i,\Gamma^i)$ are the (0,2) components of the (2,2) multiplet $X_i$.  Without loss of generality we can restrict attention to $0 < q_i < 1/2$.\footnote{We assume $q_i >0$.  In that case for a non-singular potential any fields with $q_i \ge 1/2$ can be eliminated by their equations of motion.}  A special case occurs when we can split the fields $\{X_i\} \to \{X_a\} \cup \{X_p\}$ so that $W = W^1(X_a) + W^2(X_p)$.   This leads to an enhanced symmetry, but the enhancement is very large:  on both the left and right we obtain two $N=2$ superconformal algebras with $\cb_1$ and $\cb_2$ that add up to the total $\cb$.  The enhanced right-moving $\GU(1)$ symmetry is not part of an N=2 Kac-Moody algebra: there are two commuting N=2 superconformal algebras, and each $\GU(1)$ is the lowest component of a different $N=2$ algebra.  Thinking of this theory as a (0,2) LG model and carrying out  $c$-extremization leads to the same result for $\cb$ and charges of the chiral fields.

We will now show that in non-singular (2,2) theories this is the only way that enhanced symmetries occur.  Hence, there are no (2,2) accidents.

A necessary and sufficient condition to be able to perform the split $\{X_i\} \to \{X_a\} \cup \{X_p\}$ and $W = W^1(X_a)+W^2(X_p)$ is that the matrix of second derivatives $W_{ki}$ is block diagonal in the two sets of variables.~\footnote{We use the shorthand $W_i = \p W/\p X_i$,~ $W_{ki} = \p W /\p X_k\p X_i$~, etc.}  Since we understand the symmetry enhancement in that case, we assume that $W_{ki}$ has no non-trivial block.\footnote{Take the $n\times n$ matrix $W_{ki}$ and set to $1$ all non-zero components.  The result is a symmetric matrix $A_{ki}$ that is the adjacency matrix for a graph $G$ on $n$ nodes, with each $A_{ki} \neq 0$ specifying a path in the graph from node $k$ to node $i$.  The statement that there is no non-trivial block is simply that $G$ is connected.}
We will now show that no additional symmetry is possible when $W$ is non-singular.  The argument uses three facts.\\[2mm]
\noindent 1. A non-singular $W$ can satisfy at most one linearly independent quasi-homogeneous relation.  To see this, suppose the contrary.  By taking linear combinations of two relations we arrive at $\sum_i \alpha_i X_i W_i = 0$.  We can now split the fields $X_i$ according to $\alpha_i >0$, $\alpha_i <0$, or $\alpha_i = 0$:  $\{X_i\} \to \{Y_a\} \cup \{Z_s\} \cup \{U_\alpha\}$ and recast the relation as 
\begin{align}
\sum_a \beta_a Y_a W_a = \sum_s \gamma_s Z_s W_s~,
\end{align}
where $\beta_a,\gamma_m >0$.  Without loss of generality we may assume $\beta_1 =1 \ge \beta_a$ for $a \neq 1$.  Every monomial in $W$ that contains $Y_1$ must contain at least one $Z$.  Hence, $W$ will be singular unless $W \supset Y_1^{m} Z_s$ for some $s$, say $s=1$.  Similarly, $dW|_{Y=0}$ will be independent of $Z_1$ unless $W \supset Z_1^{p} Y_a$ for some $a$, which requires $\beta_a = \gamma_1 p = pm >1$, where the last inequality follows since $W$ has no quadratic terms in the fields.  That is in contradiction with $\beta_a \le 1$, so the theory must be singular.\\[2mm]
\noindent 2.  Suppose we have a symmetry of the (2,2) theory that commutes with the (0,2) SUSY algebra.  This means that there are charges $Q'_i$ and $q'_i$ such that 
\begin{align*}
-Q'_i W_i & = \sum_{j} q'_j X_j W_{ij}  & \implies && -Q'_i W_{ik} &= q'_k W_{ki} + \sum_j q_j X_j W_{jik}~.
\end{align*}
Exchanging $i$ and $k$ in the second equation and taking the difference, we obtain
\begin{align*}
(Q'_k -q'_k) W_{ki} = W_{ki} (Q'_i - q'_i)~.
\end{align*}
This means that whenever $W_{ki} \neq 0$ we need $Q'_k - q'_k = Q'_i - q'_i$.\footnote{This is trivially satisfied for the usual (2,2) $\GUL$, where $Q_i = q_i-1$.}\\
\noindent 3. The (2,2) superpotential satisfies
\begin{align*}
(q'_i -Q'_i) W -\sum_j q'_j X_j W_j =  U^i~,
\end{align*}
where $U^i$ is independent of $X^i$.  This follows by integrating the quasi-homogeneity condition obeyed by $W_i$.

Using these observations, we now complete the argument as follows.  Since $W_{ki} $ does not contain a non-trivial block, we see from the second fact that for all $k,i$ $Q'_k-q'_k  = Q'_i - q'_i$.  Combining this with the third fact, we find that $W$ satisfies a quasi-homogeneity relation $W(t^{q'_j} X_j) = t^{q'_i-Q'_i} W(X)$; the first fact then implies that either $q'_i = c q_i$ and $q'_i - Q'_i = c$, or $W$ is singular.  

\subsection{Subtleties for heterotic vacua}
We have seen that the identification of UV parameters with a deformation space of an IR CFT, while reasonably well understood for (2,2) theories, is  more subtle for (0,2) theories.  The difference is that while in  non-singular (2,2) theories enhanced symmetries are always associated to a decomposition of the UV theory into non-interacting components, this is not the case for (0,2) models.  An enhanced symmetry of a (0,2) model does generically mix with the naive $\GUR$, so that the enhanced symmetry point has a different central charge from what one might expect naively.  As illustrated by the example in section~\ref{ss:devilmodel}, the RG fixed points of a (0,2) model need not realize any CFT with the naive central charge.  

There are situations where the consequences are more benign: there is a choice of UV parameters that leads to a CFT with the expected IR symmetries, but even then the identification of UV parameters with marginal deformations of the IR theory requires a careful study of the field redefinition orbits on the space of UV parameters.  The following familiar example illustrates the issue.

\subsubsection*{An $\SO(10)$ heterotic Landau-Ginzburg orbifold}
Consider a (0,2) theory with the following field content and charge assignment
\begin{align}
\xymatrix@R=0.0mm@C=5mm{
~ 	& \theta		&\Phi_{1,2}	&\Phi_{3,\ldots, 6}	&\Gamma^{1,\ldots,7} \\
\bq	& 0			&\ff{2}{5}		&\ff{1}{5}			&-\ff{4}{5} \\
\bqb	& 1			&\ff{2}{5}		&\ff{2}{5}			&\ff{1}{5}	
}
\end{align}
It is easy to see that this symmetry leads to $r=4$ and $\cb = 9$.  The orbifold of this theory by $e^{2\pi i J_0}$ is a candidate for an internal SCFT of an $\SO(10)$ heterotic vacuum.  As described in~\cite{Distler:1993mk} that does seem to be the case: the massless spectrum is organized into sensible $\SO(10)$ multiplets, and there is a reasonable large radius interpretation in terms of a rank $4$ holomorphic bundle on a complete intersection CY manifold in $\C\P^5_{111122}$.  The generic superpotential for this theory is 
\begin{align}
\cW = \sum_A \Gamma^A J_A~,
\end{align}
where each $J_A$ has charge $\bq = 4/5$.  We can choose the UV parameters of the theory to produce the following non-singular potential:
\begin{align}
\cW_1 = \Gamma^1 \Phi_1^2 + \Gamma^2 \Phi_2^2 + \sum_{i=3}^6 \Gamma^i \Phi_i^4 + \Gamma^7 \times 0~.
\end{align}
This is a product of (2,2) minimal models and a free left-moving fermion.  The resulting central charge is $\cb =3 (3 + \ff{1}{15})$.  Thus, this choice of UV parameters does not correspond to a point in the moduli space of the $\cb = 9$ CFTs.
Of course the orbit of field redefinitions of this point yields a large basin of attraction of UV theories that flow to the same CFT with $\cb = 3(3+\ff{1}{15})$.  In this case we can identify another point that does lead to $\cb = 9$:  
\begin{align}
\cW_2 = \Gamma^1 \Phi_1^2 + \Gamma^2 \Phi_2^2 + \sum_{i=3}^6 \Gamma^i \Phi_i^4+ \Gamma^7 \Phi_1\Phi_2 ~.
\end{align}
While this superpotential still has a $\GU(1)^6$ global symmetry, $c$-extremization leads to $\cb = 9$ and R-charges as in the table above.  Clearly there is a relevant deformation by $\Gamma^7\Phi_1\Phi_2$ that leads to an RG flow from the $\cb = 3(3+\ff{1}{15})$ theory to the $\cb = 9$ CFT.

The general lesson is clear: field redefinitions stratify the space of UV parameters into orbits, and in general these orbits correspond to different IR fixed points that are not related by marginal deformations --- in particular they can have different central charges.  The orbits may or may not include an IR fixed point for which  the manifest symmetry of the generic superpotential becomes the $\GURIR$ :  in this example they do, while in that of section~\ref{ss:devilmodel} they do not.

\section{Marginal deformations of a unitary (2,0) SCFT} \label{s:CPT}
This section contains a number of results on (2,0) SCFTs.  Many if not all of these are well-known in the context of heterotic compactifications, but the derivations given here are more general and give a useful alternative perspective.

\subsection{Basic results}
Consider a unitary compact (2,0) SCFT  with the usual superconformal algebra generators $J(z)$, $G^\pm(z)$, and $T(z)$, with modes given respectively by $J_n$, $G^\pm_r$ and $L_n$.\footnote{While for many purposes it is very convenient to treat the supersymmetric side of the theory as anti-holomorphic, in the discussion that follows it leads to a great profusion of bars.  Hence, in this section the SUSY side will be taken to be holomorphic.}

We will show that marginal Lorentz-invariant and supersymmetric deformations of this theory by a local operator take the form
\begin{align}
\label{eq:marclaim}
\Delta S &= \int \!\! d^2 z ~\Delta \cL~,&  \Delta \cL &= \AC{G^-_{-1/2}}{\cU} + \text{h.c.}~,
\end{align}
where $\cU$ is a chiral primary operator with $\GUL$-charge $\bq = 1$ and weights $(h,\hb) =(1/2,1)$.\footnote{Some of the arguments given here were developed by IVM and MRP in collaboration with Ido Adam.}   In string theory, where one considers (0,2) SCFTs with quantized $\bq$ charges, this is a classic result~\cite{Dixon:1987bg}.  Here we will apply the point of view developed for $N=1$ $d=4$ SCFTs~\cite{Green:2010da} to arrive at the statement without any assumptions of charge integrality.  

\subsection*{Constraints from supersymmetry}
Without loss of generality we can consider deformations $\delta \cL = \cO(z,\zb)$ by a quasi-Virasoro primary operator $\cO$, since a descendant would just be a total derivative.  Lorentz invariance requires $\cO$ to have spin $0$, i.e. $h_{\cO} = \hb_{\cO}$.  In order for $\delta \cL$ to be supersymmetric, we need $\CO{G^\mp_{-1/2}}{\cO}$ to be a total derivative, i.e. $G^\mp_{-1/2} |\cO\ra = {L}_{-1}|\cM^\mp\ra$.  Applying $G^\pm_{-1/2}$ to both sides of the equation and using the N=2 algebra, we obtain
\begin{align}
G^{\pm}_{-1/2} G^{\mp}_{-1/2} |\cO\ra &= L_{-1} G^{\pm}_{-1/2} |\cM^{\mp}\ra~,\nonumber\\
 \implies& L_{-1} \left[ |\cO\ra - G^+_{-1/2} |\cM^-\ra -G^-_{-1/2} |\cM^+\ra \right]  = 0~.
\end{align}
Hence, up to a constant multiple of the identity operator, which does not lead to a deformation of the theory,
we can write $|\cO\ra$ as
\begin{align}
|\cO\ra = G^-_{-1/2} |\cM^+\ra + G^+_{-1/2} |\cM^-\ra ~,
\end{align}
and hence, without loss of generality, any non-trivial deformation  corresponds to a state
\begin{align}
|\cO\ra = G^-_{-1/2} |\cU\ra + G^+_{-1/2} |\cV\ra + \left[ G^+_{-1/2} G^-_{-1/2} -(1+\ff{\bq_{\cK}}{2h_{\cK}} )L_{-1}\right] |\cK\ra~,
\end{align}
where $|\cU\ra$, $|\cV\ra$ and $|\cK\ra$ are all quasi-primary with respect to the N=2 superconformal algebra, i.e. annihilated by the lowering modes of the global N=2 algebra, $L_1$ and $G^\pm_{1/2}$.  The linear combination of operators in the last term is fixed by $L_{1} |\cO\ra =0$.   The spins of the fields are
\begin{align}
\hb_{\cU} -h_{\cU} & = 1/2~,&
\hb_{\cV}-h_{\cV}  & = 1/2~,&
\hb_{\cK}-h_{\cK}  & = 1~.&
\end{align}
The remaining constraints from supersymmetry are
\begin{align}
G^+_{-1/2} G^-_{-1/2} |\cU \ra &= L_{-1} |X \ra~,&
G^-_{-1/2} G^+_{-1/2} |\cV \ra &= L_{-1} |Y \ra
\end{align}
for some states $|X\ra$ and $|Y\ra$.  We will now show that the only solution to these equations is that $\cU$ ($\cV$) is a chiral primary (anti-chiral primary ) state.  It suffices to work out the constraint on $|\cU\ra$ --- the one on $|\cV\ra$ follows by exchanging $G^+$ and $G^-$.

Without loss of generality we decompose
\begin{align}
|X \ra = a |\cU\ra + |\chi\ra~,
\end{align}
where $a$ is real and $|\chi\ra$ is orthogonal to $|\cU\ra$.  The condition now becomes
\begin{align}
(G^+_{-1/2} G^-_{-1/2} -a L_{-1}) |\cU\ra = L_{-1} |\chi\ra~.
\end{align}
Applying $\la \cU | L_{1}$ to both sides and using orthogonality of $|\cU\ra$ and $|\chi\ra$, we find $a = 1 + \ff{\bq_{\cU}}{2h_{\cU}}$.
Application of $\la \chi |L_1$ to both sides shows $L_{-1} |\chi\ra = 0$, so we are left with
\begin{align}
G^+_{-1/2} G^-_{-1/2} |\cU \ra &= (1+ \ff{\bq_{\cU}}{2h_{\cU}}) L_{-1} |\cU \ra~.
\end{align}
Finally, applying $G^-_{-1/2}$, we find 
\begin{align}
(1+\ff{1}{2h_{\cU}}) (2h_{\cU}-\bq_{\cU}) = 0~.
\end{align}
The only solution of this equation consistent with unitarity is $2h_{\cU} = \bq_{\cU}$, i.e. $|\cU\ra$ is a chiral primary state of the N=2 superconformal algebra.  

Combining the preceding results and applying them to deformations by real operators, we conclude that real Lorentz-invariant supersymmetric deformations take the form
\begin{align}
\cO(z,\zb) = \left[ \AC{G^-_{-1/2}}{\cU(z,\zb)}  +\text{h.c.} \right]+ \AC{G^+_{-1/2}}{\CO{G^-_{-1/2}}{\cK(z,\zb)}}~,
\end{align}
where $\cU$ is a fermionic chiral primary operator with $\hb_{\cU} = \ff{1}{2} + h_{\cU}$, $h_{\cU} = \bq_{\cU}/{2}$, and $\cK$ is a real bosonic quasi-primary operator with $\hb_{\cK} = 1 + h_{\cK}$.  As in four dimensions~\cite{Green:2010da}, we recognize the familiar superpotential and K\"ahler deformations.

\subsubsection*{Marginal operators}
If we impose in addition that the perturbation is marginal, we obtain the constraints $\bq_{\cU} = 1$ and $h_{\cK} =0$.  The latter implies that $L_{-1} |\cK\ra =0$, i.e. $\cK(\zb)$ is an anti-holomorphic conserved current that leads to a trivial deformation of the action.  We arrive at the result~(\ref{eq:marclaim}).

\subsection{A few consequences}
The preceding analysis, when combined with some basic assumptions about superconformal perturbation theory, leads to important constraints on (2,0) SCFTs.  The key feature is that we can use a (2,0) superspace to recast the marginal deformations into the form
\begin{align}
\Delta S =  \int d^2z  \int d \theta ~ \alpha^i \cU_i  + \text{h.c.}~,
\end{align}
where $\alpha^i$ denote the couplings and $\cU_i$  are denote the chiral primary marginal fermi superfields.  Assuming there exists a manifestly supersymmetric regularization scheme for conformal perturbation theory, the renormalized action at a renormalization scale $\mu$ must take the form
\begin{align} \Delta \cL_{\text{ren}} &= \int d^2\theta \left[ 
Z^a(\alpha,\alphab; \mu) \Jb_a  + \sum_A \mu^{2-d_A} \Kb_A\right]  \nonumber\\
~&~~~~
+ \left\{ \int d\theta \left[ (\alpha_i+\delta\alpha^i(\alpha;\mu)) \cU_i + \zeta^I(\alpha;\mu) \cU_I \right] + \text{h.c.} \right\}.
\end{align}
At the conformal point ($\alpha = 0$) the $\Jb_a$ and $\Kb_A$ are real operators of dimension $\Delta_a =2$ and $\Delta_A >2$,
while the $\cU_i$ and $\cU_I$ are chiral primary operators with $\bq=1$ and $\bq>1$ respectively.\footnote{Compactness of the CFT ensures a gap in dimensions between $\Jb_a$ and $\Kb_A$, as well as between $\cU_i$ and $\cU_I$. } 
The first line is parallels the $N=1$ $d=4$ situation; however, the second line is new, following from the
fact that the $\cU_i \cU_j$ OPE will in general have singular $\zb$ dependence.  Of course supersymmetry
still requires that the renormalization of the superpotential should be holomorphic in the parameters.  

\subsubsection*{Marginal irrelevance}
Marginal deformations preserve the R-symmetry of the original SCFT.  Hence, the unitarity bound $h_{\cU} \ge \bq /2$ implies that a marginal deformation is at worst marginally irrelevant and never marginally relevant. 

\subsubsection*{D-terms and F-terms}
Assuming that conformal perturbation theory is renormalizable, the terms involving
the $\Kb_A$ and $\cU_I$ do not arise, and scale invariance of the theory is  equivalent to 
\begin{align}
 D^a(\alpha,\alphab) \equiv \mu \frac{\p}{\p \mu} Z^a = 0 \qquad\text{and}\qquad
    \Fb^i (\alpha) \equiv \mu \frac{\p}{\p\mu} \delta \alpha^i = 0. 
\end{align}
A two-dimensional unitary compact scale-invariant theory is automatically conformal~\cite{Polchinski:1987dy}, so every deformation satisfying these ``D-term'' and ``F-term'' constraints is exactly marginal.   

The ``D-term'' obstructions to marginality are exactly the same as in the $d=4$ case studied in~\cite{Green:2010da} ---  such a scale dependence requires the breaking of a global right-moving symmetry.  This is easy to understand at leading order in conformal perturbation theory.  In the presence of abelian currents $\Jb_a$, the OPE of $\cU$ with its conjugate takes the form
\begin{align}
 \cU_i(z,\zb) \overline{\cU}_{\jb}(w,\wb) \sim \frac{g_{i\jb}}{(z-w)(\zb-\wb)^2} +\frac{ g_{i\jb} q^i_a \Jb^a(w,\wb)}{(z-w)(\zb-\wb)}~ +\ldots,
\end{align}
where $\Jb^a = \gamma^{ab} \Jb_b$, and $z\zb\la \Jb_a(\zb) \Jb_b(0)\ra = \gamma_{ab}$ in the undeformed theory.
This leads to a logarithmic divergence in conformal perturbation theory proportional to 
\begin{align}
\int d^2w G^+_{-1/2} G^-_{-1/2}~~\int d^2 z\frac{ |\alpha^i|^2 q^i_a \Jb^a(w,\wb)}{(z-w)(\zb-\wb)} \sim \log\mu \times |\alpha^i|^2 q^i_a \Jb^a(w,\wb)~,
\end{align}
which corresponds to the leading order D-term proportional to
\begin{align}
D_a = \sum_i |\alpha^i|^2 q^i_a~.
\end{align}
 In applications to heterotic
compactifications such a symmetry necessarily corresponds to a gauge boson in the space-time theory,
and the space-time picture of the breaking is just the Higgs mechanism:  the obstruction to marginality of a coupling $\alpha$ that breaks a right-moving symmetry is encoded in a space-time D-term potential.  
%

We believe that in (0,2) LG models without an orbifold there are no F-term obstructions.  The reason is simple:   the free field UV presentation of the theory comes with the usual non-renormalization theorems for the superpotential, so the only divergences we expect to encounter will correspond to D-term counter-terms.

The two sources of obstruction are in one to one correspondence with the two ways in which a short chiral primary multiplet can combine into a long multiplet of (2,0) SUSY.  Suppose we consider an infinitesimal (2,0) SUSY deformation under which a marginal chiral primary state $|\cU\ra$ acquires weights $(h,\hb) = (\ff{1}{2} + \ff{\ep}{2}, 1+ \ff{\ep}{2})$.  In this case $|\cU\ra$ is no longer chiral primary, and by a choice of basis we can consider two separate cases:
\begin{align}
G^+_{-1/2} |\cU\ra &\neq 0~,&
G^-_{1/2} |\cU\ra &=0~,& &&\text{or} &&
G^+_{-1/2} |\cU\ra &= 0~,&
G^-_{1/2} |\cU\ra \neq 0~.
\end{align}
In other words, $|\cU\ra$ remains primary but is no longer chiral, or it remains chiral but fails to be primary.  The first case corresponds to an F-term obstruction, where at $\ep=0$ we have two chiral primary superfields $(\cU,\cF)$ with $\bq_{\cU} =1$ and $\bq_{\cF}=2$ and $\hb =1$, while for $\ep>0$ we find a complex long multiplet with lowest component $|\cU\ra$ and $G^+_{-1/2}|\cU\ra = \sqrt{\ep} |\cF\ra$.\footnote{In a $c=9$ theory with spectral flow there is a canonical $\cF$ for every $\cU$ in the theory.  Indeed, as observed in~\cite{Atick:1987gy,Dine:1987gj,Dixon:1987bg}, the $\cF$ (2,0) superfields can be used to construct vertex operators for the space-time auxiliary fields residing in chiral multiplets of the associated four-dimensional theory.} The second case corresponds to a D-term obstruction, where at $\ep =0$ we have chiral primary superfields $\cU$, its anti-chiral conjugate $\overline{\cU}$, and a Kac-Moody current $\Jb$; for $\ep >0$ we obtain a long real multiplet with lowest component $\Jb$ and descendants $G^+_{-1/2} |\Jb\ra =\sqrt{\ep} |\cU\ra$ and $G^-_{-1/2} |\Jb\ra = \sqrt{\ep} |\overline{\cU}\ra$.

In particular, we see that there are no F-term obstructions if the undeformed theory has no chiral primary operators with $\bq=2$ and $\hb =1$.  This is the case, for instance, in every (2,0) SCFT with $c<6$.  If there are also no left-moving Kac-Moody symmetries then every (2,0) marginal deformation must remain exactly marginal.   In appendix~\ref{app:Fterm} we mention a simple example illustrating an F-term obstruction at $c=9$.

\subsubsection*{K\"ahler geometry of the moduli space}
One can use the same reasoning as in~\cite{Green:2010da} to argue that the space of truly marginal deformations of a (2,0) SCFT must be a K\"ahler manifold.  This is because the D-term constraints and the quotient by global symmetries lead to a toric quotient on the space of marginal couplings, while the F-term constraints are manifestly holomorphic constraints, restricting the truly marginal directions to a K\"ahler subvariety of the toric variety.  In heterotic compactification this can of course be argued either from the space-time heterotic supergravity or by using additional assumptions of a (2,0) SCFT with integral charges~\cite{Periwal:1989mx}.  The argument given here is more direct and general.

\subsubsection*{Application to (2,2) theories}
The case of a (2,2) SCFT and its (2,2)-preserving deformations is much simpler.  There are two types
of superpotential deformations: the chiral and the twisted chiral.  The former corresponds to deformations by chiral primary (c,c) ring operators, while the latter by the (a,c) ring operators.  Supersymmetry implies that twisted
chiral parameters can never show up in the renormalized chiral superpotential and vice-versa.  Moreover, 
the OPE of the (c,c) and (a,c) chiral primaries with themselves is non-singular, so that neither potential is corrected---there are no F-term obstructions to marginality.  Hence, all marginal (c,c) and (a,c) symmetry-preserving deformations are truly marginal.  This is again a familiar story in string applications~\cite{Dine:1986zy,Dixon:1987bg}~.

\subsubsection*{Accidents beyond field redefinitions}
We can now see that the field redefinitions of (0,2) LG theories do not describe all accidents.  It is not the case
that every direction transverse to field redefinition orbits corresponds to a marginal deformation of the IR theory.
This is due to the possibility that marginal deformations of a (0,2) theory can turn out to be marginally irrelevant.   In (0,2) LG theories this is due to D-term obstructions where a $\GU(1)$ symmetry is broken by turning on operators with a definite sign of the $\GU(1)$ charge.  We give an example of this phenomenon in a well-known heterotic vacuum in appendix~\ref{app:Dterm}.

\subsection{Deformations and left-moving abelian currents}
As a final application of the preceding results, we consider the interplay between deformations of a (2,0) SCFT and left-moving currents.

A (2,0) SCFT may possess a KM algebra on the SUSY side of the world-sheet in addition to the $\GUL$ current $J_L$ in the $N=2$ multiplet.  Such structures are familiar from heterotic compactifications preserving $8$ space-time supercharges in four dimensions --- when realized geometrically these correspond to geometries $\pi : X \to K3$ --- principal $T^2$ fibrations over a base $K3$~\cite{Becker:2006et,Melnikov:2012cv}.  In each such case we can use a Sugawara-like decomposition to decompose the $N=2$ world-sheet superconformal algebra (SCA) into two commuting sets of generators, one associated to the KM algebra, and the other corresponding to the remaining degrees of freedom.  

Suppose we have an abelian current algebra $\GU(1)$ with current $J_1$.  There are two ways that the decomposition can work.  If $J$ does not belong to a multiplet of the $\cA_{c}$ SCA, then we must have a decomposition
\begin{align}
\cA_{c} = \cA'_{c'} \oplus \cA''_{c''}~,
\end{align}
where $c = c'+c''$, and the lowest components of the N=2 multiplets of $\cA'$ and $\cA''$ are obtained by appropriate linear combinations of $J_L$ and $J_1$.  We are familiar with such examples from above:  this happens whenever the LG theory decomposes into a product of two non-interacting theories.

If $J$ does belong to a multiplet of $\cA$, then it must be accompanied by a second $\GU(1)$ Kac-Moody current $J_2$, as well as weight $h=1/2$ operators $\psi_1$ and $\psi_2$.  Together these arrange themselves into a well-known  $c=3$ unitary representation of $N=2$:
\begin{align}
\label{eq:c3}
J_L  & =  :\bpsi \bpsib:~,&
G^+ & =  \sqrt{2} \bpsi \bjb~, &
G^-  & =  \sqrt{2}\bpsib \bj~,\quad ~,&
T & =  :\bj\bjb: -\ff{1}{2} (:\bpsib\p\bpsi:+:\bpsi\p\bpsib:)~,
\end{align}
where $\bpsi$ and $\bj$ have the free-field OPEs
\begin{align}
\bpsib(z) \bpsi(w) \sim (z-w)^{-1},\qquad
\bjb(z) \bj(w) \sim  (z-w)^{-2}.
\end{align}
This is equivalent to the holomorphic sector of a $T^2$ (1,0) non-linear sigma model, and we will call it $\cA^{\text{free}}_3$.

There is a key difference between these two generalizations.  In the first case, there are generally deformations that can break the extra left-moving symmetry --- in (2,2) LG this happens when we move away from a Gepner point to a more generic theory.  In the second case such breaking is impossible.  To see this, we just need to apply what we learned about the structure of SUSY deformations in conformal perturbation theory.  Since our algebra splits as
\begin{align}
\cA_{c} = \cA'_{c-3} \oplus \cA^{\text{free}}_{3}~,
\end{align}
a marginal deformation has a similar decomposition
\begin{align}
\cU = \cU' + S \bpsi~,
\end{align}
where $\cU'$ is a chiral primary operator with $\hb =1$ and $\bq' = 1$, while $S$ is a (0,1) current.  The deformation of the action is then
\begin{align}
G^-_{-1/2} \cdot \cU = G'^{-}_{-1/2} \cdot \cU' + \sqrt{2} S \bj~.
\end{align}
This is neutral with respect to $J_L$, $\bj$ and $\bjb$.   More generally, any relevant deformation must be of the form $\cU = \cU'$ with $\bq' < 1$.

\section{Toric geometry of the deformation space} \label{s:toric}
In the previous sections we saw that accidental symmetries play an important role in (0,2) Landau-Ginzburg theories, and more generally, in (0,2) SCFTs.  In this section we will describe a conjecture that allows us to account for these accidents in a certain class of (0,2) LG theories.  In that context our goal is to describe the moduli space $\cal M$ of IR fixed points corresponding
to a class of UV data determined by a choice of charges $q_i$ and
$Q_A$ which have the expected central charge 
\begin{align}
\cb &= 3(n-N+r)~,&  r = -\sum_A Q_A -\sum_i q_i~.
\end{align} 
To do so, we need to perform two steps:
\begin{enumerate}
\item decompose the UV parameter space into orbits under the action of
  field redefinitions;
\item determine which orbits contribute to $\cal M$.
\end{enumerate}
The result is expected to be a (typically singular) K\" ahler space.  In general these are rather formidable tasks. 
The group of field redefinitions is rather large and the space of orbits is non-separable.  A reasonable geometry can only emerge
after implementing the second task.  This involves excluding two types of orbits:
\begin{itemize}
\item Along a \textit{discriminant locus} $\Delta$ in parameter space, the
  superpotential is singular.   The discriminant will clearly be
  invariant under field redefinitions, and  orbits contained in $\Delta$
  will not contribute to $\cM$.
\item For some non-singular values of the parameters, the theory will
  have accidental symmetries in the IR.  As we have seen, in some
  cases these symmetries will mix nontrivially with the R-symmetry and
  the central charge of the IR fixed point will be larger than $\cb$.
  Thus, these orbits as well need to be excluded from $\cal M$.
\end{itemize}
In general, the second step is difficult even if one restricts
attention to symmetries which act diagonally on the UV fields.  Detecting the 
basin of attraction of some component of the IR moduli space with
central charge $\cb'>\cb$ requires a determination of the R-symmetry
along each such component to find which deformations away from
this locus are in fact irrelevant.

\subsection{The toric conjecture}
There is a simpler version of both of these problems that may be
tractable.  The group of field redefinitions always contains an
abelian subgroup, the complexification of the $U(1)^n\times U(1)^N$ subgroup
of the global symmetry of the free kinetic terms,  
that corresponds to rescaling the chiral fields of the
theory.  In particular, if we write the most general superpotential in our class as
\begin{align}
\cW = \sum_A \Gamma^A \sum_{m\in \Delta_A} \alpha_{Am} \prod_i \Phi_i^{m_i}~,
\end{align}
where
\begin{align}
\Delta_A = \{ m \in \Z^{n} ~~|~~ \textstyle\sum_i m_i q_i = -Q_A \}~
\end{align}
describes the lattice points in the Newton polytope for $J_A$, then the field redefinitions
\begin{align}
\label{eq:plain}
\Phi_i &\mapsto t_i \Phi_i~,  &
\Gamma^A & \mapsto \tau_A \Gamma^A~
\end{align}
lead to a $T_\C = (\C^\ast)^{N-n+1}$ action\footnote{The rank of the $\C^\ast$ action is reduced by
  $1$ due to the quasi-homogeneity of $\cW$.}
 on the space of UV parameters
$Y = \C^{\sum_A |\Delta_A|}$\begin{align} 
\alpha_{Am} \mapsto \tau_A \textstyle\prod_i
t_i^{m_i} \times \alpha_{Am}~.
\end{align}  We will refer to these
as toric field redefinitions.   

We now restrict attention to these toric actions in \textit{both} of the tasks listed above.
Namely, we decompose the parameter
space into $T_\C$ orbits and exclude those orbits that either lie in
$\Delta$ or exhibit accidental symmetries contained in $T_\C$ and lead to $\cb' > \cb$.  The result, which we will call
$\cM_T$, will in some cases be equivalent to $\cM$, but in general the two will differ.  We will
comment on this further below.

The action of the compact torus $T\subset T_\C$~, given by restricting to $|t|=|\tau|=1$~,
determines a moment map $\mu=(\lambda;\Lambda):Y\to \R^{N+n}$ with
\begin{align}
\lambda_i &= \sum_{A} \sum_{m\in \Delta_A} m_i |\alpha_{Am}|^2~,&
\Lambda_A & = \sum_{m\in \Delta_A} |\alpha_{Am}|^2~.
\end{align}
Quasi-homogeneity of $\cW$ implies that the image lies in the
hyperplane 
\begin{align}
\sum_i q_i \lambda_i +\sum_A Q_A \Lambda_A = 0~.
\end{align}
The image of $\mu$ is the intersection of this hyperplane with a cone,
determined by the charges, inside the positive orthant in $\R^{N+n}$.
This intersection is itself a cone $\widehat\Sigma$, of dimension
$N+n-1$.   The level sets of $\mu$ determine a selection of orbits:
generic orbits will be $N+n-1$ dimensional, but the action will degenerate
along points with a non-trivial stabilizer subgroup, leading to
orbits of smaller dimension.\footnote{There are orbifold singularities
  when the subgroup is discrete; we will focus on continuous
  stabilizer subgroups.}  More precisely, the cone $\widehat\Sigma$ can be
subdivided into a fan $\Sigma$, such that the collection of orbits
containing a point for which $\mu(\alpha) =\mu^\ast$ is determined by the
cone of $\Sigma$ that contains $\mu^\ast$.  This is the {\sl
secondary fan\/} for the $T$ action.   We now have enough structure
to state our conjecture.\\[0mm]

\subsubsection*{Conjecture}
The toric moduli space ${\cal M}_T$ is the complement of the discriminant subvariety $\Delta$ in a {\sl
  toric variety\/}
\begin{align}
V = \mu^{-1}(\lambda^\ast,\Lambda^\ast)/T~,
\end{align}
where 
\begin{align}\label{eq:special}
\lambda^\ast_i = 1-q_i~,\qquad \Lambda_A^\ast = 1 + Q_A\ .
\end{align}
This is a rather strong statement, and we will not provide a complete proof but rather some evidence for it.  Some
ideas on a possible derivation are discussed in section~\ref{ss:bs}.  We will motivate the conjecture by combining
our results and observations from above with some facts
about toric varieties. 

We should note a few important points.  First, $V$ may turn out to be empty.  Second, while we claim that $V \setminus \Delta$ describes $\cM_T$ as a variety, we do not make any statement about the relation between the Zamolodchikov metric on the space of marginal couplings and the metric on $V$ obtained by the K\"ahler quotient.  Finally, in this paper we will be concerned with orbits of continuous field redefinitions.  In general there will be additional discrete quotients that identify points in $\cM_T$.

\subsubsection*{Combinatorics of the secondary fan}
Codimension-one cones in $\Sigma$ are
associated with orbits containing a point at which a single
$\C^\ast\subset T_\C$ is unbroken.  More precisely, $G_{(q',Q')}=\C^\ast \subset T_\C$
acting with charges ${q'_i,Q'_A}$ on the chiral
superfields will fix points at which
\begin{align}
|\alpha_{Am}|^2 \left( Q'_A + \textstyle{\sum_i} m_i q'_i \right) = 0~\qquad \text{for all}~~A,~m\in\Delta_A~.
\end{align}
The $\mu$-image of the $T_\C$ orbits of such points will lie in a cone $\sigma_{(q',Q')}$ generated
by the charge vectors of the $\alpha_{Am}$ fixed by $G_{(q',Q')}$.
Thus, the codimension-one cones of $\Sigma$ are determined by
one-dimensional subgroups for which $\sigma_{(q',Q')}$ has dimension
$N+n-2$ and lies in the hyperplane
\begin{align}\label{eq:hyperplane}
\sum_A Q'_A\Lambda_A + \sum_i q'_i\lambda_i = 0~.
\end{align}
In terms of $\cW$ the codimension-one cones of $\Sigma$ correspond to
subgroups for which we can write a (possibly singular) family of
models fixed precisely by $(\C^\ast)^2$.  Cones of higher codimension
in the fan are boundaries of these cones and
arise at the intersections of these hyperplanes.\footnote{Note that
  this does not imply that cones of higher codimension correspond to
  models with larger unbroken symmetry: values of $\mu$ at the
  intersection of two codimension-one cones can be in the image of two
  distinct $T_\C$ orbits, each of which is fixed by a different
  subgroup.} Points in the
interior of some cone of the fan (of any codimension) lie in the image
of a collection of orbits determined by that cone.  Our conjecture is thus
equivalent to the statement that the $\mu$-image of the $T_\C$ orbits of models with central
charge $\cb$  intersects the cone $\sigma^\ast$ containing $\mu^\ast$ in its interior.

A cone $\sigma\in\Sigma$ can be specified by its relation to the
codimension-one cones $\sigma_{(q',Q')}$.  For each of these, $\sigma$
either lies inside $\sigma_{(q',Q')}$, in which case $\mu\in\sigma$
satisfy~(\ref{eq:hyperplane}), or it lies on one side or the other, meaning~(\ref{eq:hyperplane}) is satisfied as a strict
inequality for all $\mu\in\sigma$.  To prove our claim we thus need to
show that orbits of points in parameter space corresponding to models
with central charge $\cb$ are precisely those containing in their
image points satisfying the inequalities satisfied by $\mu^\ast$.
To do this we must consider all codimension-one cones of $\Sigma$.
We classify these by the nature of the models exhibiting the enhanced
symmetry. 

\subsection{Enhanced toric symmetries}

\subsubsection*{Symmetries realized by a non-singular potential}
Consider first the case of one-parameter subgroups of $T_\C$ for
which the generic point in the locus they fix corresponds to a
nonsingular model with a $\GU(1)^2$ global symmetry.  
The IR R-symmetry can then be determined
by $\cb$-extremization as 
\begin{align}
\hat q_i &= t q_i + s q'_i~,&\hat Q_A = t Q_A + s Q'_A~,
\end{align}
where
\begin{align}\label{eq:extremtwo}
L \begin{pmatrix} t-1\\s\end{pmatrix} = 
\begin{pmatrix} 0\\\sum_i q'_i\lambda^\ast_i + \sum_A
  Q'_A\Lambda^\ast_A\end{pmatrix}~.
\end{align}
 $L$ is the negative-definite $2\times 2$ matrix defined in~(\ref{eq:knl}) and $(q,Q)$ are normalized as in~(\ref{eq:normfix}).
We now distinguish two situations.\\[2mm]   

\noindent\underline{1.  $\cb' = \cb$}. If
\begin{align}\label{eq:nojump}
\sum_i q'_i\lambda^\ast_i + \sum_A  Q'_A\Lambda^\ast_A = 0~,
\end{align}
then the IR symmetry is given by $(q;Q)$, and the $T_\C$ orbit
of the model with enhanced symmetry is a point in $V$.  In this case,
we can apply conformal perturbation theory to deformations of this
theory.  The symmetry-breaking couplings $\alpha_{Am}$ (those
vanishing on the locus exhibiting enhanced symmetry) parameterize
classically marginal deformations away from the symmetric theory. 
The analysis of section 4 shows that in fact some of these will be
marginally irrelevant, and the moduli space is given to first order in
the symmetry-breaking couplings by the vanishing of the D-term for
the broken symmetry.   We can write this explicitly here as
\begin{align}
D &= \sum_{A,n\in\Delta_A} (Q'_A + \sum_i m_i q'_i)|\alpha_{Am}|^2\nonumber\\
&= \sum_i q'_i\lambda_i + \sum_A Q'_A\Lambda_A~.
\end{align}
This holds at leading order in conformal perturbation theory about the symmetric point, and our conjecture
amounts here to the statement that higher order corrections do not qualitatively modify the structure of the
symplectic quotient that leads to the variety $V$:  while the metric may be modified, which orbits are kept and which are excluded is not changed by higher order corrections. This  implies
that points in $V$ are $T_\C$ orbits containing points whose 
image under $\mu$ lies in the cone $\sigma_{(q',Q')}$.  We see from~(\ref{eq:nojump}) that this
condition is satisfied by $\mu^\ast$.\\[2mm]

\noindent\underline{2. $\cb' > \cb$.}  If~(\ref{eq:nojump}) is not satisfied, the central charge $\cb'$ determined
by extremization will be larger than $\cb$,
and the $T_\C$ orbit of the model with enhanced symmetry is not a point of $V$.  Moreover, the symmetry-breaking parameters
$\alpha_{Am}$ are not marginal couplings in this theory.
Solving~(\ref{eq:extremtwo}) we find 
\begin{align}
s = -{r\over {\rm det} L}\left(\sum_i q'_i\lambda^\ast_i + \sum_A
  Q'_A\Lambda^\ast_A\right)~.
\end{align}
Without loss of generality we can choose the sign of $(q',Q')$ so that $s$ is negative.
Since by construction all our couplings are invariant under $(q;Q)$, and by assumption $L$ is negative definite,
the sign of the charge under the IR R-symmetry is then the opposite of the charge under
$(q',Q')$.  Thus, couplings $\alpha_{Am}$ for which 
$Q'_A + \sum_i m_i q'_i>0$ will be {\sl relevant\/} deformations of
the model with enhanced symmetry, while couplings with the opposite charge
will be irrelevant; couplings preserving the enhanced symmetry are marginal.  The $T_\C$ orbits of points in parameter space
corresponding to irrelevant deformations of the symmetric model will
not be points in $V$: as discussed above they will exhibit an
accidental symmetry in the IR and a central charge $\cb'$.  Orbits
for which at least one relevant coupling is nonzero are
characterized precisely by the fact that they contain points for which
the moment map satisfies
\begin{align}
\sum_i q'_i\lambda_i + \sum_A Q'_A\Lambda_A > 0~.
\end{align}
This specifies one side of the hyperplane associated to the enhanced
symmetry, and, as we have shown, this is the side on
which the point $\mu^\ast$ lies.

\subsubsection*{Symmetries without a smooth realization}

If every enhanced symmetry were realized by a non-singular $\cW$ the discussion
above would suffice.  In general, however, there are 
codimension-one cones in $\Sigma$ associated
to one-parameter subgroups of $T_\C$ for which it is not possible to
construct a non-singular $\cW$ exhibiting the symmetry.   
In these
cases the RG trajectories exhibiting the enhanced symmetry along the
flow are singular, and we cannot use their properties to determine the
local structure of the moduli space. 

A simple example of this is
given by the symmetry acting as $\Gamma^A\to \tau\Gamma^A$ for some
$A$ with all the other fields invariant.  This fixes the locus $J_A=0$ which
will in general be singular (it will always be singular when $n=N$).
In this case, the corresponding hyperplane is $\Lambda_A=0$, and the associated
codimension-one cone lies on the boundary of $\Sigmah$.

More interesting is the case of a codimension-one cone in the interior of
$\widehat\Sigma$ to which the methods of the previous section do not apply.
Our conjecture here is that whenever the enhanced symmetry does not
satisfy~(\ref{eq:nojump}), non-singular models will only exist when at 
least one symmetry-breaking coupling whose charge under the broken 
symmetry is in accord with the sense of the inequality is non-zero.  
In parallel with the second discussion in the previous subsection, 
$T_\C$ orbits associated to points in $V$ will be those containing 
points whose image under $\mu$ lies on the side of the 
hyperplane which contains the point $\mu^\ast$.

  There will also be codimension-one cones in $\Sigma$ associated to
  one-parameter subgroups for which there is no non-singular model
  exhibiting the symmetry, but which satisfy~(\ref{eq:nojump}).  Here
  as well we can classify the symmetry-breaking couplings by their
  charge under the broken symmetry.  In this case, we conjecture that 
  non-singular models will have nonzero values for at
  least one coupling of {\sl each\/} sign.  Restricting to models with
  non-zero couplings of only one sign (as well as the neutral
  couplings) will produce a singular model.  The space of $T_\C$ orbits
  associated to points in $V$ in this case will not be toric.  It can,
  however, be described as the complement of the symmetric locus (a
  component of $\Delta$) in a (singular) toric variety.  This
  contains orbits containing points whose image under $\mu$ lies {\sl
    in\/} the cone $\sigma_{(q',Q')}$.  When we exclude the singular
    symmetric locus here, we find precisely orbits that have nonzero
    symmetry-breaking couplings with both signs of the broken charge.
The point $\mu^\ast$ clearly lies in this hyperplane.

\subsection{Examples}

A few examples may be helpful at this point.  We proceed from a simple example for which our methods produce correctly the actual moduli space to models demonstrating their limitations.

\subsubsection*{A plain model}
Consider first the class of models with $n=N=2$ and charges
\begin{align}
\xymatrix@R=2mm@C=0mm{
~ 	& \Phi_1		&\Phi_2		&\Gamma^1		&\Gamma^2	\\
\bq	& \ff{1}{4}	&\ff{1}{6}	& -\ff{3}{4}		&-\ff{5}{6}			}
\end{align}
and $\cb/3 = r = 1 + \ff{1}{6}$.
The most general superpotential is 
\begin{align}\label{eq:easy}
\cW = \Gamma^1\left(\alpha_{11}\phi_1^3 + \alpha_{12}\phi_1\phi_2^3\right)
 + \Gamma^2\left(\alpha_{21}\phi_2^5 +
   \alpha_{22}\phi_1^2\phi_2^2\right)~,
\end{align}
and the discriminant is 
\begin{align}\label{eq:discex}
\Delta =
\alpha_{11}\alpha_{21}\left(\alpha_{11}\alpha_{21}-\alpha_{12}\alpha_{22}\right)~.
\end{align}
This is an example of what we call a {\sl plain\/} model: the torus
$T_\C$ includes all field redefinitions consistent with the symmetry,
so our toric considerations will in fact generate the moduli space
$\cal M$ itself.

The torus $T=\GU(1)^3$ action on $\C^4$ is characterized by the charges and moment map components
\begin{align}
\xymatrix@R=2mm@C=0mm{
  D 	          & \alpha_{11}		& \alpha_{12}	& \alpha_{21}	& \alpha_{22}	\\
  \lambda_1	& 3                &1	                        &0			&2\\
  \lambda_2	& 0                 &3	                       &5			&2\\
  \Lambda_1	& 1                 &1	                       &0			&0\\
  \Lambda_2	&  0                &0	                        &1			&1
}
\end{align}
where the latter satisfy 
\begin{align}
3\lambda_1 + 2\lambda_2 = 9\Lambda_1 + 10\Lambda_2~ .
\end{align}
There are six codimension-one cones in $\Sigma$.  Only two of these are
realized by non-singular models; the remaining four comprise the boundaries of $\widehat\Sigma$ given by
 $\Lambda_1>0$~ and $\Lambda_2>0$, as well as $\lambda_2-2\Lambda_2>0$ and $\lambda_1-\Lambda_1>0$.  

There are two codimension-one cones in the interior of $\widehat\Sigma$.  Consider first
the symmetry $(q';Q') = (1,0;-3,0)$, which satisfies~(\ref{eq:nojump}).  The 
non-singular models realizing this symmetry have $\alpha_{12} = \alpha_{22} = 0$.  In fact the model
reduces to a product of two (2,2) minimal models and, as expected, the central charge 
is $\cb$.   The symmetry determined by $(q',Q') = (-1,1;-2,0)$, for which 
$\sum_i q'_i\lambda^\ast_i + \sum_A  Q'_A\Lambda^\ast_A < 0$, fixes models with $\alpha_{11}=\alpha_{22}=0$.\footnote{Note that this symmetry leads a $2\times 2$ L matrix that is not negative-definite; however, the corresponding superpotential is singular.}  Under the broken symmetry, $\alpha_{11}$ is negatively charged
and $\alpha_{22}$ positively charged.  
We see from~(\ref{eq:discex}) that, in accordance with the conjecture, non-singular models require a non-zero value for the negatively charged coupling.

The moduli space $\cal M$ is thus determined.  We can fix two of the generators of $T_\C$ by setting $\alpha_{11}=\alpha_{21}=1$, and the remaining couplings parameterize the toric variety $V=\C$ with invariant coordinate $z = \alpha_{12}\alpha_{22}$.  The moduli space is ${\cal M} = V\setminus\widetilde\Delta$ where the discriminant reduces in these coordinates to $1-z$.

\subsubsection*{A non-plain model}

We can also consider the model with $n=N=2$ and charges
given by 
\begin{align}
\xymatrix@R=2mm@C=0mm{
~ 	& \Phi_1		&\Phi_2		&\Gamma^1		&\Gamma^2	\\
\bq	& \ff{46}{471}	&\ff{115}{471}	& -\ff{460}{471}		&-\ff{230}{471}			}
\end{align}
with $\cb/3 = r = 1 + \ff{58}{471}$.  The most general superpotential invariant under this symmetry is 
\begin{align}
\cW = \Gamma^1 (\alpha_{11} \Phi_1^{10} + \alpha_{12} \Phi_1^5\Phi_2^2 + \alpha_{13} \Phi_2^4) + \Gamma^2 (\alpha_{21} \Phi_1^5 + \alpha_{22}\Phi_2^2)~,
\end{align}
and the discriminant is 
\begin{align}
\Delta = \alpha_{11} \alpha_{22}^2-\alpha_{12}\alpha_{21}\alpha_{22}+\alpha_{13}\alpha_{21}^2~.
\end{align}

The torus $T = \GU(1)^3$ action on $\C^5$ is characterized by the charges and moment map components
\begin{align}
\xymatrix@R=2mm@C=0mm{
  D 	          & \alpha_{11}		& \alpha_{12}		& \alpha_{13}	& \alpha_{21}	& \alpha_{22}	\\
  \lambda_1	& 10                &5	                        &0		        &5			&0\\
  \lambda_2	& 0                 &2	                        &4		        &0			&2\\
  \Lambda_1	& 1                 &1	                        &1		        &0			&0\\
  \Lambda_2	&  0                &0	                        &0		        &1			&1
}
\end{align}
where the latter satisfy 
\begin{align}
2\lambda_1 + 5\lambda_2 = 20\Lambda_1 + 10\Lambda_2\ .
\end{align}
The cone $\widehat\Sigma$ is the intersection of this with the positive orthant.
This is bounded, in this case, by the coordinate hyperplanes.  There are four codimension-one 
cones in the interior of $\widehat\Sigma$ here, none of which satisfy~(\ref{eq:nojump}). 

The symmetries acting with charges $(q';Q') = (1,0;-5,-5)$ and $(q',Q') = (1,0,-5,0)$ are preserved 
by singular models, and non-singular models, as per the conjecture, lie in orbits containing points for which $\lambda_1-5\Lambda_1-5\Lambda_2<0<\lambda_1-5\Lambda_1$.  The symmetry acting with charges $(q',Q') = (1,0;-10,0)$ fixes the locus $\alpha_{12} =
\alpha_{13}=\alpha_{21}=0$ where we find a product of (2,2) minimal models: up to
a rescaling
\begin{align}
\cW_1 = \Gamma^1\Phi_1^{10} +\Gamma^2 \Phi_2^2~,
\end{align}
with central charge $\cb_1/3 = 1 + \ff{5}{33}>r$.   At this point, the operators
associated to $\alpha_{12}$ and $\alpha_{13}$ are {\sl irrelevant} but
the operator associated to $\alpha_{21}$ is {\sl relevant}.  We
conclude that models with $\alpha_{11}\alpha_{22}\neq 0$ and
$\alpha_{21}=0$ flow to this IR fixed point and orbits containing such 
models do not contribute to $V$.  Orbits that do contribute have a point for which 
$\lambda_1 > 10\Lambda_1$.

The symmetry acting with charges $(q',Q') = (0,1;-4,0)$ fixes the locus $\alpha_{11} = \alpha_{12} = \alpha_{22}=0$ where we find a product of (2,2) minimal models: up to
a rescaling 
\begin{align}
\cW_2 = \Gamma^1\Phi_2^{4} +\Gamma^2 \Phi_1^5~,
\end{align}
with central charge $\cb_2/3 = 1 + \ff{4}{15}>\cb_1/3$.    At this point, the operators
associated to $\alpha_{11}$ and $\alpha_{12}$ are {\sl irrelevant} but
the operator associated to $\alpha_{22}$ is {\sl relevant}.  We
conclude that models with $\alpha_{13}\alpha_{21}\neq 0$ and
$\alpha_{22}=0$ flow to this IR fixed point and orbits containing such 
models do not contribute to $V$.  The orbits that do contribute have a point for
which $\lambda_2 > 4\Lambda_1$.

Our toric model $V$ of the moduli space is thus determined here by the cone
\begin{align}
\lambda_1 > 10 \Lambda_1~,\qquad \lambda_2 > 4 \Lambda_1~,\qquad \Lambda_1>0~.
\end{align}
Applying~(\ref{eq:special}) we find that, as expected, the point 
\begin{align}
\mu^* = (\ff{435}{471},\ff{356}{471};\ff{11}{471},\ff{241}{471})~
\end{align}
lies in this cone.
Points in the preimage of this have $\alpha_{21}$ and $\alpha_{22}$ 
both non-zero.    We
can use two of our rescalings to fix $\alpha_{21} = \alpha_{22}=1$,
and under the remaining symmetry the three coefficients in $J_1$
transform homogeneously, so we have $V = \P^2$.  Of course, this is an
overparametrization.   This is not a plain model, and we can use the remaining
field redefinitions $\Gamma^2\to\Gamma^2 + \Gamma^1(a\Phi_1^5 + b\Phi_2^2)$ to 
show that these theories flow to a unique IR fixed
point.  Not unrelated to this is the fact that there is no
discriminant here: any point in $\P^2$ corresponds to a non-singular
model.  

\subsubsection*{A model with $N>n$}

The model discussed in section~\ref{ss:devilmodel} shows more of the limitations of toric methods.  
Here we have $Y = \C^9$ and $T_\C = (\C^\ast)^4$ acts on the couplings.   $\widehat\Sigma$ is the intersection of $\lambda_1+3\lambda_2 = 6(\Lambda_1 + \Lambda_2+\Lambda_3)$ with the positive orthant.  There are a total of 18 codimension-one cones in the interior of $\widehat\Sigma$.   Proceeding with our method we find a five-dimensional toric variety $V$ determined by the moment map values 
$\mu^\ast = (\ff67,\ff47;\ff17,\ff17,\ff17)$.  This is a puzzle, since we found previously that there are no models in this class with $\cb=3$.  The resolution is that the model
\begin{align}
\cW'_3 = \Gamma^1 \Phi_1^6 + \Gamma^2 \Phi_2^2 + \Gamma^3\Phi_2^2~,
\end{align}
which is in the inverse image under $\mu$ of the point $\mu^\ast$, is fixed by a $\GU(1)$ rotation in the $\Gamma^{2,3}$ plane that is not contained in $T_\C$.  This is a symmetry which arises as an accidental symmetry for all points in $V$, and is manifest for $\cW'_3$.  This mixes with the IR R-symmetry leading to the central charge found above.  This phenomenon in which a Fermi field is in fact free in the IR can occur in non-plain models with $N>n$.  For models with $N=n$ a model with a free Fermi field will be singular.

\subsection{Summary and further thoughts}\label{ss:bs}

We have provided evidence for a strong conjecture on the structure of the 
space of $T_\C$ orbits contributing to $V$.  For models in which these are 
the only field redefinitions consistent with the UV symmetry this produces 
the moduli space $\cal M$ of SCFTs with central charge $\cb$.  
By analogy with studies of (0,2) GLSM parameter
spaces~\cite{Kreuzer:2010ph,Melnikov:2010sa}, we call these {\sl plain\/} models. 
For models with
larger groups of field redefinitions, our discussion is partial in two
ways: we have overparametrized the moduli space, and we have failed,
in general, to exclude the basins of attraction of models in which a
symmetry in the complement of $T_\C$ mixes with the IR R-symmetry.

Our evidence, while suggestive, falls short of a derivation of 
the result.  The key difficulties in a proof are twofold.  First, the consequences of enhanced symmetries that are only realized by singular superpotentials are difficult to grasp, since we do not have conformal perturbation theory as a guide.  For these our evidence is based on the analysis of many examples that all turned out to be consistent with the conjecture.  The second difficulty lies in extending the leading order conformal perturbation theory result for enhanced symmetry loci with $\cb' =\cb$.   It may be possible to improve this by a more detailed study of the 
combinatorial structures involved.

  A more satisfactory derivation can be imagined, 
which proceeds by constructing a $\cb$ function along the RG flow and showing that
this can be written in terms of $\alpha$ through the combinations forming $(\lambda;\Lambda)$, 
along the lines of~\cite{Kutasov:2003ux,Kutasov:2004xu,Erkal:2010sh}.\footnote{A conversation with D. Kutasov,
in which he suggested this idea, was instrumental in leading us to the results of this section.}  
In that work, global symmetries broken by couplings were incorporated into $a$-maximization in four-dimensional theories by imposing constraints on the space of symmetries over which one 
maximized a trial $a$-function.  The Lagrange multipliers implementing the constraints could then be used to parameterize the flow.  In our case the symmetry-breaking couplings are the superpotential couplings which break the global symmetry $U(1)^{N+n}$ of the (free) UV theory to $U(1)$ and one can
introduce Lagrange multipliers to constrain the symmetries over which $\cb$ is extremized.  Of course,
imposing $N+n-1$ constraints is a formal procedure, because this is tantamount to specifying the outcome.  However, if one proceeds formally, one finds an expression for $\cb$ in terms of the Lagrange multipliers and the {\sl values\/} of these at the extremum --- which reproduces~(\ref{eq:boringcharges}) --- are precisely the values of the moment map given by~(\ref{eq:special}).   The relation between this
formal result and the values of the moment map is not clear to us.

\section{Outlook}
This project began as an attempt to classify IR fixed points of (0,2) LG theories --- a generalization of the results obtained for (2,2) LG theories in~\cite{Kreuzer:1992np,Klemm:1992bx}~.  This beautiful work shows that for fixed $c=\cb$ there is a finite set of families of superpotentials $W(X_1,\ldots,X_n)$, or equivalently charges $\bq(X_i)$ that lead to a non-singular (2,2) SCFT of desired central charge.  Having the (0,2) generalization would be very useful:  we would have a new class of heterotic vacua and more generally (0,2) SCFTs with many properties computable in terms of the simple UV description.  These would naturally fit into the class of (0,2) gauged linear sigma models and could be used to produce a large class of hybrid models along the lines of~\cite{Bertolini:2013xga}.

What we learned is that, in contrast to the (2,2) case, it is not enough to classify non-singular (0,2) potentials realizing a particular set of $\GUL\times\GUR$ charges.  For instance, the model studied section~\ref{ss:devilmodel} would naively realize a $c=4$, $\cb=3$ (0,2) SCFT that could correspond to some rather exotic $8$-dimensional heterotic vacuum.  In fact no such IR fixed point is obtained for any choice of the UV parameters.  This is a general lesson for building UV models of (0,2) SCFTs:  a check of UV R-symmetry anomalies is not enough, and while the UV theory may well flow somewhere (i.e to an SCFT with $\cb >0$), it may wind up far (i.e. at infinite distance) from the expectations of the model builder.  We expect this to be a general lesson applicable to the wider class of gauged linear sigma models.  In exploring that latter point it should be interesting to study in detail GLSMs with LG phases that exhibit accidents and extrapolate their consequences to large radius geometries.

For a class of models---the plain LG theories---we were able to obtain a compelling conjecture for a global description of the (0,2) moduli space $\cM$ realizing the expected central charge.  While the resulting combinatorial structure is consistent with a case-by-case analysis of field redefinitions and their orbits in examples, we were not able to prove it in generality.  Progress on both testing and proving the conjecture could be made by developing a better understanding of the combinatorial structure of quasi-homogeneous (0,2) superpotentials, as well as developing Lagrange multiplier techniques and trial $\cb$ functions.  A classification of plain LG theories seems achievable; this would yield a large playground to explore LG RG flows and could give hints to the more general classification problem.  Finally, it should be illuminating to relate our work to studies of RG flows with redundant couplings, e.g.~\cite{Behr:2013vta}.

\appendix
\section{An F-term obstruction} \label{app:Fterm}
In this section we give an example, taken from~\cite{Aspinwall:2010ve}, that illustrates both D-term and F-term obstructions to marginal couplings.  The setting is a (2,2) LG orbifold (LGO) compactification of the heterotic string with a superpotential
\begin{align}
W =  X_0^4 + X_1^4 + X_2^4 + X_3^8 + X_4^8 + \psi X_0 X_1 X_2 X_3 X_4 + \ep \Delta W~.
\end{align}
Here $\psi$ and $\ep$ are parameters and $\Delta W$ is a generic polynomial with $\bq = 1$.
Marginal (2,0) deformations of the LGO correspond to massless $\GE_6$-neutral space-time chiral multiplets.  We can compute the massless spectrum exactly as a function of the complex parameters in the superpotential using the technique developed in~\cite{Kachru:1993pg}. This leads to the following results.
\begin{enumerate}
\item Setting $\psi = \ep = 0$ leads to a $\GU(1)^4$ right-moving Kac-Moody algebra and $298$ marginal (2,0) deformations.  We now turn on the (2,2)-preserving $\psi$ and $\ep$ deformations and investigate what happens to the remaining (2,0) deformations.  From above we know that at worst the marginal (2,0) deformations can become marginally irrelevant. 
\item With $\psi \neq 0$ but $\ep =0$ the $\GU(1)^4$ symmetry is broken, and the number of marginal (2,0) deformations is $298-4-6 = 288$.  While $4$ of the $10$ marginally irrelevant deformations are associated to the broken symmetries the $6$ others are not.
\item Finally, turning  on $\ep \neq 0$ does not break any continuous symmetries, but the number of marginal (2,0) deformations decreases to $288-6 = 282$.
\end{enumerate}
Note however, that all the singlets lifted by F-terms correspond to twisted sectors of the LGO.  This is consistent with there being no F-term obstructions in pure LG theories.

\section{A D-term obstruction in a heterotic vacuum} \label{app:Dterm}
Consider now the (2,2) quintic LG theory coupled to a free left-moving fermion with
\begin{align}
\cW = \sum_{i=1}^5 \Gamma^i J_i(\Phi) =  \sum_{i=1}^5 \Gamma^i (\Phi_i^4 + \psi \prod_{j\neq i} \Phi_j) + \Gamma^6\times 0~.
\end{align}
The interacting fields have their usual $\GUL$ charges $q_i = 1/5$ and $Q_i = -4/5$, and for $\psi \neq 0$ there are no extra $\GU(1)$ symmetries in addition to $\GUL\times\GUR\times\GU(1)_6$, where $\GU(1)_6$ is the symmetry associated to the free $\Gamma^6$.  This flows to a conformal field theory with $r = 4$ and $\cb = 9$, and we can now consider deformations of the IR theory from the general perspective of deforming by chiral primary operators.  In the (2,2) theory we have a good understanding of the map between the IR chiral primary marginal operators and the UV data, so
we can identify the marginal deformations of the IR theory with the space of possible $\cW$ modulo field redefinitions.   If we keep $\Gamma^6$ free, we find a $301$--dimensional space of deformations.  

We can also include deformations of the form $\Gamma^6 J_6$, where $J_6$ is some generic degree $5$ polynomial.  Although these break the $\GU(1)_6$ symmetry, they preserve the central charge and the $\GUL\times\GUR$ quantum numbers of the fields.  In particular, $\Gamma^6$ has the quantum numbers of a free field.  Including these $J_6$ deformations yields a $402$-dimensional space of marginal deformations away from the (2,2) $r=4$ $\cb = 9$ fixed point.  

Are all of these $402$ marginal deformations exactly marginal?  While all of the $301$ 
deformations of the $J_i$ are truly marginal, the $101$ extra deformations associated to $J_6\neq 0$ are marginally irrelevant.  This is completely clear from the conformal perturbation theory discussion we gave in the text.  All of these break the $\GU(1)_6$ symmetry, and every symmetry-breaking coupling has the same sign of $\GU(1)_6$ charge.  Let us now see how the same result is recovered from a  heterotic space-time perspective.

\subsubsection*{Heterotic insights}
The $\Z_5$ orbifold of the LG theory just described, combined with an appropriate heterotic GSO projection leads to a well-understood heterotic vacuum:  the LG point in the moduli space of the quintic compactification with standard embedding.  The massless fields of the resulting space-time N=1 $d=4$ supergravity theory consist of the supergravity multiplet, the axio-dilaton chiral multiplet, the $\Le_6\oplus\Le_8$ vector multiplets, $326$ gauge-neutral chiral multiplets, and a $\Le_6$ charged chiral spectrum $\rep{27}\oplus\brep{27}^{\oplus 101}$.  The $301$ deformations of the $J_i$ described above correspond to $\Le_6$-preserving marginal deformations in the untwisted sector of the orbifold.  These remain truly marginal for any value of the K\"ahler modulus (itself in a twisted sector), and at large radius they are the $101$ complex structure deformations of the CY quintic, as well as $200$ of the $224$ deformations of the tangent bundle.  As reviewed in~\cite{McOrist:2011bn}, there are many arguments for why these deformations are truly marginal.

The $101$ deformations associated with the $J_6$ couplings also have a simple space-time interpretation: they correspond to $\so(10)$-singlet components of the $\brep{27}^{\oplus 101}$.  Turning on these deformations corresponds to Higgsing $\Le_6\to \so(10)$.  This makes it obvious that the deformations are marginally irrelevant.  Under the decomposition of $\Le_6 \supset \so(10) \oplus \Lu(1)$, we have 
\begin{align}
\brep{27} = \brep{16}_{-1/2} \oplus\rep{10}_{1} \oplus \rep{1}_{-2}~.
\end{align}
The $\so(10)$ singlets all have charge $-2$ under the broken $\Lu(1)$, and hence have a D-term space-time potential.
This is an example of a ``D-term'' obstruction to a marginal coupling being exactly marginal.

Since the deformation only involves world-sheet fields in untwisted sector of the orbifold, it is clear by the orbifold inheritance principle that this obstruction to marginality lifts to the un-orbifolded quintic LG model and matches the conformal perturbation theory result.   In the orbifold theory it is possible to find exactly marginal deformations that Higgs $\Le_6\to \so(10)$~\cite{Dine:1988kq}, but they involve an interplay between marginal couplings in twisted and untwisted sectors~\cite{McOrist:2011bn}.

\bibliographystyle{./utphys}
\bibliography{./bigref}

\end{document}